\documentclass[12pt,aps,prd,showpacs,amsmath,amssymb]{revtex4}
\input epsf
\begin{document}
\title{Light-cone Distribution Amplitudes of $\Xi$ and their Applications}
\author{Yong-Lu Liu and Ming-Qiu Huang}
\affiliation{Department of Physics, National University of Defense Technology, Hunan 410073, China}
\date{\today}
\begin{abstract}
We present the light-cone distribution amplitudes of the $\Xi$ baryons up to twist six on the basis of QCD conformal partial wave expansion to the leading order
conformal spin accuracy. The nonperturbative parameters relevant to the DAs are determined in the framework of the QCD sum rule. The light-cone QCD sum rule
approach is used to investigate both the electromagnetic form factors of $\Xi$ and the exclusive semileptonic decay of $\Xi_c$ as applications. Our estimations on
the magnetic moments are $\mu_{\Xi^0}=-(1.92\pm0.34)\mu_N$ and $\mu_{\Xi^-}=-(1.19\pm0.03)\mu_N$. The decay width of the process $\Xi_c\rightarrow \Xi e^+\nu_e$ is
evaluated to be $\Gamma=8.73\times10^{-14}\; \mbox{GeV}$, which is in accordance with the experimental measurements and other theoretical approaches.
\end{abstract}
 \maketitle

\section{Introduction}
\label{sec1} Light-cone distribution amplitudes (DAs), which denote the momentum fraction distributions of partons in a hadron, play important roles in hard
exclusive processes \cite{exclusive,exclusive2} and the light-cone QCD sum rule applications. Furthermore, DAs are of eminent importance themselves because of their
ability to reveal the internal structure of the composite particles. Studies on the nucleon octet DAs were first carried out early in late 1980s with QCD sum rules
on the moments up to the leading twist \cite{Chernyak}. More than ten years later, Braun $et$ $al.$ gave a systematic investigation on the higher order DAs of the
nucleon based on the conformal expansion \cite{Braun1}. Recently, an overview was given in Ref. \cite{overviewofDAs}, in which the authors present a comparison of
the nucleon DAs with various approaches and models. In our previous work \cite{DAs}, we have given the DAs of the $\Sigma$ and $\Lambda$ baryons on the conformal
partial wave expansion to the leading order conformal spin accuracy. As composite particles with two $s$-quarks, $\Xi$ have the same Lorentz structures as other
$J^P=\frac{1}{2}^+$ octet baryons in the $SU(3)$-flavor symmetry limit, but the $SU(3)$-flavor breaking effects may play more important roles than in the cases of
the other octet baryons, provided that the masses of the $s$-quarks are considered. Thus investigations on the $\Xi$ baryon DAs are of interest and somewhat
complicated.

The higher order twist contributions to DAs have several origins, among which the main contribution comes from ``bad'' components in the wave function and in
particular of components with ``wrong'' spin projection for the case of baryons \cite{Braun1,DAs}. We focus on higher order twist contributions from ``bad''
components in the decomposition of the Lorentz structure in this paper. The general description of DAs is based on the conformal symmetry of the massless QCD
Lagrangian dominated on the light cone. The conformal partial wave expansion of the DAs can be carried out safely in the limit of the $SU(3)$-flavor symmetry
approach. However, when terms connected with the $s$-quark mass are considered, the $SU(3)$-flavor breaking effects need to be included. We assume that the
conformal expansion of the nucleon DAs can be extrapolated to our cases, which is similar to the arguments for mesons and baryons with a single $s$-quark
\cite{DAs,mesondas}. In the present work, effects from the $SU(3)$-flavor symmetry breaking are considered as the corrections, which originate from two sources:
isospin symmetry breaking and corrections to the noperturbative parameters.

The related processes containing the baryons can be investigated in the framework of the light-cone QCD sum rule (LCSR)\cite{lcsr1,lcsr2,lcsr3} with the DAs. LCSR
is a developed noperturbative method of the traditional QCD sum rule \cite{SVZ}. Its main idea is to expand the correlation function between the vacuum and the
hadron state on the light cone $x^2=0$, while the noperturbative effects, which correspond to the condensates in traditional QCD sum rules, are described by the DAs
connected with the final hadron state. LCSR has been widely used to investigate the electromagnetic (EM) form factors and various decay processes related to the
baryons \cite{DAs,LYL,Lenz,Aliev,Wzg,Huang,Wang,Wym}. In the applications, we take advantage of the LCSR approach to investigating the EM form factors of $\Xi$ and
form factors of the weak transition $\Xi_c\rightarrow\Xi$. Both the magnetic moments of $\Xi$ baryons and decay width of the semileptonic transition
$\Xi_c\rightarrow \Xi e^+\nu_e$ will be estimated.

The paper is organized as follows. Section \ref{sec:def} is devoted to give the definitions of the DAs and necessary parameters. In Sec. \ref{sec:conexp}, the
conformal partial wave expansion of the DAs is carried out by use of the conformal symmetry. The nonperturbative parameters connected with the DAs are determined in
Sec. \ref{sec:sumrule}. Section \ref{sec:app} is the application part, in which the EM form factors of $\Xi$ and the semileptonic decay of $\Xi_c$ baryons are
investigated in the framework of LCSR. The summary and conclusion are given in Sec. \ref{sec:sum}. Finally, we give the explicit expressions of the $\Xi$ baryon DAs
in the Appendix.
\section{The light-cone distribution amplitudes of $\Xi$}\label{sec:def}
Our discussion of the $\Xi$ baryon DAs in this section is an extension and a complement of \cite{Braun1,DAs}, so we just list the results following their
procedures, and it is recommended that the original papers be consulted for details. Generally, DAs of the $J^P=\frac{1}{2}^+$ octet baryons can be defined by the
matrix element of the three-quark operator between the vacuum and the baryon state $|B(P)\rangle$ in the limit of $SU(3)$-flavor symmetry:
\begin{equation}
\langle{0} |\epsilon^{ijk} {q_1}_\alpha^i(a_1 z) {q_2}_\beta^j(a_2 z) {q_3}_\gamma^k(a_3 z) |{B(P)}\rangle,\label{matele}
\end{equation}
where $\alpha$, $\beta$, $\gamma$ refer to the Lorentz indices and $i$, $j$, $k$ the color indices, $q_i$ represents the light quark, $z$ is a lightlike vector
which satisfies $z^2=0$, and $a_i$ are real numbers denoting coordinates of valence quarks.

For the case of $\Xi$, after taking into account the Lorentz
covariance, spin and parity of the baryons, the matrix element
(\ref{matele}) is decomposed as
\begin{equation}
4 \langle{0} |\epsilon^{ijk} s_\alpha^i(a_1 z) s_\beta^j(a_2 z) q_\gamma^k(a_3 z) |{\Xi(P)}\rangle = \sum\limits_{i}\mathcal{F}_i\,\Gamma_{1i}^{\alpha\beta}\Big
(\Gamma_{2i}\Xi\Big )_\gamma \,,\label{da-def}
\end{equation}
where $\Xi_\gamma$ is the spinor of the baryon with the quantum number $I(J^P)=\frac12(\frac{1}{2}^+)$ ($I$ is the isospin, $J$ is the total angular momentum and
$P$ is the parity), $\Gamma_{1(2)i}$ are certain Dirac structures over which the sum is carried out, and $\mathcal{F}_i=\mathcal{S}_i, \mathcal {P}_i, \mathcal
{A}_i, \mathcal{V}_i,\mathcal{T}_i$ are the distribution amplitudes which depend on the scalar product $P\cdot z$.

As these ``calligraphic'' invariant functions do not have a definite twist in the above definition (\ref{da-def}), the twist classification is carried out in the
infinite momentum frame. With the aid of the definition of light-cone DAs with a definite twist,
\begin{equation}
4\langle {0}| \epsilon^{ijk} s_\alpha^i(a_1 z) s_\beta^j(a_2 z) q_\gamma^k(a_3 z) |{\Xi(P)}\rangle =\sum\limits_{i}F_i\,\Gamma_{1i}'^{\alpha\beta}\Big
(\Gamma_{2i}'\Xi^{\pm}\Big )_\gamma\,,\label{da-deftwist}
\end{equation}
the invariant functions $\mathcal S_i,\mathcal P_i,\mathcal V_i,\mathcal A_i,\mathcal T_i$ can be expressed in terms of the DAs $F_i=S_i,P_i,V_i,A_i,T_i$. A simple
derivation leads to the following relations between these two sets of definitions for scalar and pseudo-scalar distributions,
\begin{eqnarray}
\renewcommand{\arraystretch}{1.7}
\hspace*{-4.7cm}
\begin{array}{lll}
{\cal S}_1 = S_1\,, &\qquad\qquad\qquad& 2p\cdot z\, {\cal S}_2 = S_1-S_2\,, \\
{\cal P}_1 = P_1\,, &\qquad\qquad\qquad& 2p\cdot z\, {\cal P}_2 = P_2-P_1\,,
\end{array}
\renewcommand{\arraystretch}{1.0}
\end{eqnarray}
for vector distributions,
\begin{eqnarray}
\renewcommand{\arraystretch}{1.7}
\begin{array}{lll}
\mathcal V_1 = V_1\,, &~~& 2 p\cdot z \mathcal V_2 = V_1 - V_2 - V_3\,, \\
2 \mathcal V_3 = V_3\,, &~~& 4 p\cdot z \mathcal V_4 = - 2 V_1 + V_3 + V_4 + 2 V_5\,, \\
4 p\cdot z \mathcal V_5 = V_4 - V_3\,, &~~& (2 p\cdot z )^2 \mathcal V_6 = - V_1 + V_2 + V_3 + V_4 + V_5 - V_6\,,
\end{array}
\renewcommand{\arraystretch}{1.0}
\end{eqnarray}
for axial-vector distributions,
\begin{eqnarray}
\renewcommand{\arraystretch}{1.7}
\begin{array}{lll}
\mathcal A_1 = A_1\,, &~~& 2 p\cdot z \mathcal A_2 = - A_1 + A_2 - A_3\,, \\
2 \mathcal A_3 = A_3\,, &~~& 4 p\cdot z \mathcal A_4 = - 2 A_1 - A_3 - A_4 + 2 A_5\,, \\
4 p\cdot z \mathcal A_5 = A_3 - A_4\,, &~~& (2 p\cdot z )^2 \mathcal A_6 = A_1 - A_2 + A_3 + A_4 - A_5 + A_6\,,
\end{array}
\renewcommand{\arraystretch}{1.0}
\end{eqnarray}
and, finally, for tensor distributions,
\begin{eqnarray}
\renewcommand{\arraystretch}{1.7}
\begin{array}{lll}
\mathcal T_1 = T_1\,, &~~& 2 p\cdot z \mathcal T_2 = T_1 + T_2 - 2 T_3\,, \\
2 \mathcal T_3 = T_7\,, &~~& 2 p\cdot z \mathcal T_4 = T_1 - T_2 - 2 T_7\,, \\
2 p\cdot z \mathcal T_5 = - T_1 + T_5 + 2 T_8\,, &~~&
(2 p\cdot z)^2 \mathcal T_6 = 2 T_2 - 2 T_3 - 2 T_4 + 2 T_5 + 2 T_7 + 2 T_8\,,\\
4 p\cdot z \mathcal T_7 = T_7 - T_8\,, &~~& (2 p\cdot z)^2 \mathcal T_8 = -T_1 + T_2 + T_5 - T_6 + 2 T_7 + 2 T_8 \,.
\end{array}
\renewcommand{\arraystretch}{1.0}
\end{eqnarray}

The classification of the DAs $F_i$ with a definite twist are listed
in Table \ref{tabDA-def}. The explicit expressions of the definition
can be found in Refs. \cite{Braun1,DAs}. Each distribution amplitude
$F_i$ can be represented as
\begin{equation}
F(a_ip\cdot z)=\int \mathcal Dxe^{-ipz\sum_ix_ia_i}F(x_i),
\end{equation}
where the dimensionless variables $x_i$, which satisfy the relations
$0<x_i<1$ and $\sum_ix_i=1$, correspond to the longitudinal momentum
fractions carried by the quarks inside the baryon. The integration
measure is defined as
\begin{equation}
\int \mathcal Dx=\int_0^1dx_1dx_2dx_3\delta(x_1+x_2+x_3-1).
\end{equation}
\begin{table}
\renewcommand{\arraystretch}{1.1}
\caption{Independent baryon distribution amplitudes in the chiral expansion.}
\begin{center}
\begin{tabular}{|l|l|l|l|}
\hline& Lorentz structure & Light-cone projection & Nomenclature
\\ \hline
Twist 3 &  $ \left(C\!\not\!{z}\right) \otimes \!\not\!{z} $ &
$s^+_\uparrow s^+_\downarrow u^+_\uparrow$ & $\Phi_3(x_i) =
\left[V_1-A_1\right](x_i)$ \\ \hline&  $ \left(C i \sigma_{\perp z}
\right) \otimes \gamma^\perp \!\not\!{z} $ & $s^+_\uparrow
s^+_\uparrow u^+_\downarrow$ & $T_1(x_i)$ \\  \hline Twist 4 &  $
\left(C\!\not\!{z}\right) \otimes \!\not\!{p} $ & $s^+_\uparrow
s^+_\downarrow u^-_\uparrow$ & $\Phi_4(x_i) =
\left[V_2-A_2\right](x_i)$
\\ \hline
& $ \left(C\!\!\not\!{z}\gamma_\perp\!\!\not\!{p}\,\right)
      \otimes \gamma^\perp\!\!\not\!{z} $
& $ s^+_\uparrow s^-_\downarrow u^+_\downarrow$ & $\Psi_4(x_i) = \left[V_3-A_3\right](x_i)$ \\ \hline & $ \left(C \!\not\!{p}\!\not\!{z}\right)  \otimes
\!\not\!{z}$ & $s^-_\uparrow s^+_\uparrow  u^+_\uparrow$ & $\Xi_4(x_i) = \left[T_3-  T_7 + S_1 + P_1\right](x_i)$ \\ \hline & $ \left(C
\!\not\!{p}\!\not\!{z}\right) \otimes \!\not\!{z}$ & $s^-_\downarrow s^+_\downarrow  u^+_\uparrow$ & $\Xi_4'(x_i) = \left[T_3 + T_7 + S_1 - P_1\right](x_i)$\\
\hline& $ \left(C i \sigma_{\perp z} \right) \otimes \gamma^\perp \!\not\!{p} $ & $s^+_\downarrow s^+_\downarrow u^-_\downarrow$ & $T_2(x_i)$
\\ \hline
Twist 5 &  $ \left(C\!\not\!{p}\right) \otimes \!\not\!{z} $ &
$s^-_\uparrow s^-_\downarrow u^+_\uparrow$ & $\Phi_5(x_i) =
\left[V_5-A_5\right](x_i)$
\\ \hline
& $ \left(C\!\!\not\!{p}\gamma_\perp\!\!\not\!{z}\,\right) \otimes \gamma^\perp\!\!\not\!{p} $ & $ s^-_\uparrow s^+_\downarrow
u^-_\downarrow $ & $\Psi_5(x_i) = \left[V_4-A_4\right](x_i)$ \\
\hline & $ \left(C \!\not\!{z}\!\not\!{p}\right)  \otimes \!\not\!{p}$ & $s^+_\uparrow s^-_\uparrow  u^-_\uparrow$ &
$\Xi_5(x_i) = \left[-T_4- T_8 + S_2 + P_2\right](x_i)$ \\
\hline & $ \left(C \!\not\!{z}\!\not\!{p}\right)  \otimes \!\not\!{p}$ & $s^+_\downarrow s^-_\downarrow  u^-_\uparrow$ & $\Xi_5'(x_i) = \left[S_2 - P_2-T_4+
T_8\right](x_i)$ \\ \hline&  $ \left(C i \sigma_{\perp p} \right) \otimes \gamma^\perp \!\not\!{z} $ & $s^-_\downarrow s^-_\downarrow u^+_\downarrow$ & $T_5(x_i)$
\\ \hline
Twist 6 &  $ \left(C\!\not\!{p}\right) \otimes \!\not\!{p} $ &
$s^-_\uparrow s^-_\downarrow u^-_\uparrow$ & $\Phi_6(x_i) =
\left[V_6-A_6\right](x_i)$ \\ \hline&  $ \left(C i \sigma_{\perp p}
\right) \otimes \gamma^\perp \!\not\!{p} $ & $s^-_\uparrow
s^-_\uparrow u^-_\downarrow$ & $T_6(x_i)$ \\ \hline
\end{tabular}
\end{center} \label{tabDA-def}
\end{table}

There are some symmetry properties of the DAs from the identity of the two $s$-quarks in the $\Xi$ baryons, which are useful to reduce the number of the independent
DAs. Taking into account the Lorentz decomposition of the $\gamma$-matrix structure, it is easy to see that the vector and tensor DAs are symmetric, whereas the
scalar, pseudoscalar and axial-vector structures are antisymmetric under the interchange of the two $s$-quarks:
\begin{eqnarray}
V_i(1,2,3)&=&\;\ V_i(2,1,3),\hspace{2.8cm} T_i(1,2,3)=\;\ T_i(2,1,3),\nonumber\\
S_i(1,2,3)&=&-S_i(2,1,3),\hspace{2.7cm} P_i(1,2,3)=-P(2,1,3),\nonumber\\
A_i(1,2,3)&=&-A(2,1,3).
\end{eqnarray}
The similar relationships hold for the calligraphic structures in Eq. (\ref{da-def}).

In order to expand the DAs by the conformal partial waves, we rewrite the DAs in terms of quark fields with definite chirality
$q^{\uparrow({\downarrow})}=\frac{1}{2}(1\pm\gamma_5)q$. The classification of the DAs in this presentation can be interpreted transparently: projection on the
state with the two $s$-quarks antiparallel, i.e. $s^\uparrow s^\downarrow$, singles out vector and axial-vector structures, while parallel ones, i.e. $s^\uparrow
s^\uparrow$ and $s^\downarrow s^\downarrow$, correspond to scalar, pseudoscalar and tensor structures. The explicit expressions of the DAs by chiral-field
representations are presented in Table \ref{tabDA-def} as an example for $\Xi^0$. The counterparts of $\Xi^-$ can be easily obtained under the exchange
$u\leftrightarrow d$.

Note that in the case of the nucleon, the isospin symmetry can be used to reduce the number of the independent DAs to eight. However, there are no similar isospin
symmetric relationships that existe when the $\Xi$ baryon is considered, which is the same as the cases of the $\Lambda$ and $\Sigma$ baryons. Therefore, we need
altogether $14$ chiral-field representations to express all the DAs.

\section{Conformal expansion}\label{sec:conexp}
In this section we will give the explicit expressions of the DAs with the aid of the conformal expansion approach. The conformal expansion of the DAs is based on
the conformal symmetry of the massless QCD Lagrangian, which makes it possible to separate longitudinal degrees of freedom from transverse ones. The properties of
transverse coordinates are described by the renormalization scale that is determined by the renormalization group, while the longitudinal momentum fractions that
are living on the light cone are governed by a set of orthogonal polynomials, which form an irreducible representation of the collinear subgroup $SL(2,R)$ of the
conformal group.

The algebra of the collinear subgroup $SL(2,R)$ is determined by the following four generators:
\begin{equation}
{\bf L}_+=-i{\bf P}_+,\; {\bf L}_-=\frac{i}{2}{\bf K}_-,\; {\bf L}_0=-\frac{i}{2}({\bf D}-{\bf M}_{-+}),\; {\bf E}=i({\bf D}+{\bf M}_{-+}),
\end{equation}
where ${\bf P}_\mu$, ${\bf K}_\mu$, $\bf D$, and ${\bf M}_{\mu\nu}$ correspond to the translation, special conformal transformation, dilation and Lorentz
generators, respectively. The notations are used for a vector $A$: $A_+=A_\mu z^\mu$ and $A_-=A_\mu p^\mu/p\cdot z$. Let ${\bf L}^2={\bf L}_0^2-{\bf L}_0+{\bf
L}_+{\bf L}_-$; then a given distribution amplitude with a definite twist can be expanded by the conformal partial wave functions that are the eigenstates of ${\bf
L}^2$ and $L_0$.

For the three-quark state, the distribution amplitude with the lowest conformal spin $j_{min}=j_1+j_2+j_3$ is \cite{Braun2,Balitsky}
\begin{equation}
\Phi_{as}(x_1,x_2,x_3)=\frac{\Gamma[2j_1+2j_2+2j_3]}{\Gamma[2j_1]\Gamma[2j_2]\Gamma[2j_3]}{x_1}^{2j_1-1}{x_2}^{2j_2-1}{x_3}^{2j_3-1},\label{as-dis}
\end{equation}
where $j_i$ represents the conformal spin of the quark field. Contributions with higher conformal spin $j=j_{min}+n$ ($n=1,2,...$) are given by $\Phi_{as}$
multiplied by polynomials that are orthogonal over the weight function (\ref{as-dis}). In our approach the calculation just contains the leading order conformal
spin expansion. For DAs in Table \ref{tabDA-def}, we give their conformal expansions:
\begin{eqnarray}
\Phi_3(x_i)=120x_1x_2x_3\phi_3^0(\mu),\hspace{2.5cm} T_1(x_i)=120x_1x_2x_3\phi_3'^0(\mu),\label{contwist3}
\end{eqnarray}
for twist three and
\begin{eqnarray}
\Phi_4(x_i)&=&24x_1x_2\phi_4^0(\mu),\hspace{2.5cm}
\Psi_4(x_i)=24x_1x_3\psi_4^0(\mu),\nonumber\\
\Xi_4(x_i)&=&24x_2x_3\xi_4^0(\mu),\hspace{2.5cm}
\Xi_4'(x_i)=24x_2x_3\xi_4'^0(\mu),\nonumber\\
T_2(x_i)&=&24x_1x_2\phi_4'(\mu),\label{contwist4}
\end{eqnarray}
for twist four and
\begin{eqnarray}
\Phi_5(x_i)&=&6x_3\phi_5^0(\mu),\hspace{2.5cm}
\Psi_5(x_i)=6x_2\psi_5^0(\mu),\nonumber\\
\Xi_5(x_i)&=&6x_1\xi_5^0(\mu),\hspace{2.5cm}
\Xi_5'(x_i)=6x_1\xi_5'^0(\mu),\nonumber\\
T_5(x_i)&=&6x_3\phi_5'(\mu),\label{contwist5}
\end{eqnarray}
for twist five and
\begin{eqnarray}
\Phi_6(x_i)=2\phi_6^0(\mu),\hspace{3.5cm} T_6(x_i)=2\phi_6'(\mu), \label{contwist6}
\end{eqnarray}
for twist six. There are altogether $14$ parameters which can be determined by the equations of motion.

To the leading order, the normalization of the $\Xi$ baryon DAs is determined by the matrix element of the local three-quark operator without derivatives. The
Lorentz decomposition of the matrix element can be expressed explicitly as follows:
\begin{eqnarray}
4\langle0|\epsilon^{ijk}s^i_\alpha(0)s^j_\beta(0)q^k_\gamma(0)|\Xi(P)\rangle=\mathcal{V}^0_1(\!\not\!
PC)_{\alpha\beta}(\gamma_5\Xi)_\gamma+\mathcal{V}^0_3(\gamma_\mu
C)_{\alpha\beta}(\gamma_\mu\gamma_5\Xi)_\gamma\nonumber\\
+\mathcal{T}^0_1(P^\nu i\sigma_{\mu\nu}C)_{\alpha \beta}(\gamma^\mu\gamma_5\Xi)_\gamma+\mathcal{T}^0_3M(\sigma_{\mu\nu}C)_{\alpha
\beta}(\sigma^{\mu\nu}\gamma_5\Xi)_\gamma.
\end{eqnarray}
There are altogether four parameters to be determined. To this end, we introduce the four decay constants defined by the following matrix elements:
\begin{eqnarray}
&& \langle{0}| \epsilon^{ijk} \left[s^i(0) C \!\not\!{z} s^j(0)\right] \, \gamma_5 \!\not\!{z} q^k(0)| {P}\rangle = f_{\rm \Xi}
P\cdot z \!\not\!{z} \Xi(P)\,, \nonumber \\
&&\langle{0}| \epsilon^{ijk} \left[s^i(0) C\gamma_\mu s^j(0)\right]\, \gamma_5 \gamma^\mu q^k(0)| {P}\rangle = \lambda_1 M \Xi(P) \,,
\nonumber \\
&&\langle{0}| \epsilon^{ijk} \left[s^i(0) C\sigma_{\mu\nu} s^j(0)\right] \, \gamma_5 \sigma^{\mu\nu} q^k(0)|{P}\rangle = \lambda_2 M \Xi(P) \,,
\nonumber\\
&&\langle{0}| \epsilon^{ijk} \left[s^i(0) Ciq^\nu\sigma_{\mu\nu}s^j(0)\right]\,\gamma_5\gamma_\mu q^k(0)|{P}\rangle=\lambda_3 M\!\not\!{q}\Xi(P)
\,.\label{def-nonlocal}
\end{eqnarray}
A simple calculation gives the expressions of the local noperturbative parameters $\mathcal V_1^0, \mathcal V_3^0, \mathcal T_1^0$, and $\mathcal T_3^0$ in terms of
the four decay constants defined in Eq. (\ref{def-nonlocal}):
\begin{eqnarray}
\mathcal V_1^0&=&f_{\Xi},\hspace{3.8cm}\mathcal V_3^0=\frac
14(f_{\Xi}-\lambda_1),\nonumber\\
\mathcal T_1^0&=&\frac 16(4\lambda_3-\lambda_2),\hspace{2.1cm}\mathcal T_3^0=\frac 1{12}(2\lambda_3-\lambda_2).
\end{eqnarray}

With the above relations, the coefficients of the operators in Eqs. (\ref{contwist3})-(\ref{contwist6}) can be expressed to the leading order conformal spin
accuracy as
\begin{eqnarray}
\phi_3^0&=&\phi_6^0=f_{\Xi},\hspace{2.8cm}\psi_4^0=\psi_5^0=\frac12(f_{\Xi}-\lambda_1),\nonumber\\
\phi_4^0&=&\phi_5^0=\frac12(f_{\Xi}+\lambda_1),\hspace{1.3cm}\phi_3'^0=\phi_6'^0=-\xi_5^0=\frac16(4\lambda_3-\lambda_2),\nonumber\\
\phi_4'^0&=&\xi_4^0=\frac16(8\lambda_3-3\lambda_2),\hspace{1.2cm}\phi_5'^0=-\xi_5'^0=\frac16\lambda_2,\nonumber\\
\xi_4'^0&=&\frac16(12\lambda_3-5\lambda_2).
\end{eqnarray}
\section{QCD sum rules for the nonperturbative parameters}\label{sec:sumrule}
The nonperturbative parameters $f_{\Xi}$, $\lambda_1$, $\lambda_2$ and $\lambda_3$ of the $\Xi$ baryons can be determined in the QCD sum rule approach. The
derivations are carried out from the following two-point correlation functions,
\begin{equation}
\Pi_i(q^2)=i\int d^4x e^{iq\cdot x}\langle 0|T{j_i(x)\bar j_i(0)}|0\rangle, \label{cor}
\end{equation}
with the definitions of the currents:
\begin{eqnarray}
j_1(x)&=&\epsilon^{ijk}[s^i(x)C\!\not\! {z}s^j(x)]\gamma_5\!\not\! {z}q^k(x),\label{CZcurrent}\\
j_2(x)&=&\epsilon^{ijk}[s^i(x)C\gamma_\mu s^j(x)]\gamma_5\gamma^\mu q^k(x),\\
j_3(x)&=&\epsilon^{ijk}[s^i(x)C\sigma_{\mu\nu} s^j(x)]\gamma_5\sigma^{\mu\nu} q^k(x),\\
j_4(x)&=&\epsilon^{ijk}[s^i(x)Ciq^\nu\sigma_{\mu\nu} s^j(x)]\gamma_5\gamma^\mu q^k(x).
\end{eqnarray}
In compliance with the standard technique of the QCD sum rule, the correlation functions (\ref{cor}) need to be expressed both phenomenologically and theoretically.
By inserting a complete set of states with the same quantum numbers as those of $\Xi$, the hadronic representations of the correlation functions are given as
follows:
\begin{eqnarray}
\Pi_1(q^2)&=&2f_{\Xi}^2(q\cdot z)^3\!\not\!
{z}\frac{1}{M^2-q^2}+\int_{s_0}^\infty\frac{\rho_1^h(s)}{s-q^2}ds,\nonumber\\
\Pi_2(q^2)&=&M^2\lambda_1^2\frac{\!\not\!q+M}{M^2-q^2}+\int_{s_0}^\infty\frac{\rho_2^h(s)}{s-q^2}ds,\nonumber\\
\Pi_3(q^2)&=&M^2\lambda_2^2\frac{\!\not\!q+M}{M^2-q^2}+\int_{s_0}^\infty\frac{\rho_3^h(s)}{s-q^2}ds,\nonumber\\
\Pi_4(q^2)&=&q^2M^2\lambda_3^2\frac{\!\not\!q+M}{M^2-q^2}+\int_{s_0}^\infty\frac{\rho_4^h(s)}{s-q^2}ds.
\end{eqnarray}
On the theoretical side, we carry out the operator product expansion taking into account condensates up to dimension $6$. Then, as the usual procedure of the QCD
sum rule, we utilize the dispersion relationship and the quark-hadron duality assumption. After taking Borel transformation on both sides of the hadronic
representation and QCD expansion and matching the two sides, we arrive at the following sum rules:
\begin{eqnarray}
4(2\pi)^4f_{\Xi}^2e^{-\frac{M^2}{M_B^2}}&=&\int_{4m_s^2}^{s_0}\big\{\frac15s(1-x)^5-8m_sa_s\frac{1}{s}x^2(1-x)+\frac{1}{12}m_sm_0^2a_s\frac{1}{s^2}x(-2+3x)\nonumber\\
&&-\frac14b\frac{x(1-x)^2}{s}\big\}e^{-\frac{s}{M_B^2}}ds,\label{sumrule1}
\end{eqnarray}
and
\begin{eqnarray}
4(2\pi)^4\lambda_1^2M^2e^{-\frac{M^2}{M_B^2}}&=&\int_{4m_s^2}^{s_0}\big\{s^2[(1-x)(1-13x-x^2+x^3)-12x(1+x)\ln{x}]\nonumber\\
&&-2m_sa_s(1-x)(1+5x)-\frac13m_sa_sm_0^2\frac{1}{s}\frac{1+8x-7x^2}{1-x}\nonumber\\
&&+\frac16b(3+2x-7x^2)\big\}e^{-\frac{s}{M_B^2}}ds,
\end{eqnarray}
and
\begin{eqnarray}
(2\pi)^4\lambda_2^2M^2e^{-M^2/M_B^2}&=&\int_{4m_s^2}^{s_0}\big\{\frac32s^2[(1-x^2)(1-8x+x^2)-12x\ln{x}]-12m_sa_s(1-x^2)\nonumber\\
&&-\frac23m_sa_sm_0^2\frac{x}{s}+3b(1-x)^2\big\}e^{-\frac{s}{M_B^2}}ds,
\end{eqnarray}
and
\begin{eqnarray}
(4\pi)^4\lambda_3^2M^2e^{-M^2/M_B^2}&=&\int_{4m_s^2}^{s_0}\{s^2[(1-x)(1+17x+53x^2-11x^3)+60x^2\ln{x}]\nonumber\\
&&+\frac45s^2(1-x)^5-8m_sa_s(1-x)(1+x+4x^2)\nonumber\\
&&+2m_sa_sm_0^2\frac{1}{s(1-x)}(8-19x+12x^2-3x^3)\nonumber\\
&&-\frac{2}{3}b(1-x)(2-13x-x^2)\}e^{-\frac{s}{{M_B}^2}}ds,\label{sumrule4}
\end{eqnarray}
where the notation $x=m_s^2/s$ is adopted for convenience. In the numerical analysis, the parameters employed are the standard values: $a=-(2\pi)^2\langle\bar
uu\rangle=0.55\; \mbox{GeV}^{3}$, $b=(2\pi)^2\langle\alpha_sG^2/\pi\rangle=0.47\; \mbox{GeV}^{4}$, $a_s=-(2\pi)^2\langle\bar ss\rangle=0.8a$, $\langle\bar
ug_c\sigma\cdot Gu\rangle=m_0^2\langle\bar uu\rangle$, and $m_0^2=0.8\; \mbox{GeV}^{2}$. The mass of the strange quark is used as the central value provided by the
particle data group (PDG) \cite{PDG} $m_s=0.10\,\mbox{GeV}$.

Another important parameter in the QCD sum rule is the auxiliary Borel parameter $M_B^2$, which is introduced to suppress both higher resonance and higher order
dimension contributions simultaneously. At the same time, there should be a proper region in which the results of the sum rules vary mildly with it. The numerical
analysis shows that the working window of the Borel parameter is $0.8\; \mbox{GeV}^2\leq M_B^2\leq 1.2\; \mbox{GeV}^2$, in which our results are acceptable.

It can be seen that sum rules from Eq. (\ref{sumrule1}) to Eq. (\ref{sumrule4}) can only give the absolute values of the parameters. To determine the relative sign
of $f_{\Xi}$ and $\lambda_1$, we give the sum rule of $f_{\Xi}\lambda_1^*$:
\begin{eqnarray}
(2\pi)^4f_{\Xi}\lambda_1^*M^2e^{-\frac{M^2}{M_B^2}}&=&\int_{4m_s^2}^{s_0}\big\{-\frac12s^2x[(1-x)(2+5x-x^2)+6x\ln{x}]-\frac12m_sa_s(1-x)(3-5x)\nonumber\\
&&-\frac16m_sa_sm_0^2\frac{1+8x-7x^2}{s(1-x)}-\frac{b}{16}(1-x)(5-9x)\big\}e^{-\frac{s}{M_B^2}}ds.
\end{eqnarray}

Similarly, the relative signs of $\lambda_2$ and $\lambda_3$ to $\lambda_1$ are determined by the following two sum rules:
\begin{eqnarray}
(2\pi)^4(\lambda_1\lambda_2^*+\lambda_1^*\lambda_2)M^2e^{-\frac{M^2}{{M_B}^2}}&=&\int_{4m_s^2}^{s_0}\big\{-3m_ss^2[(1-x)(2+5x-x^2)+6x\ln{x}]\nonumber\\
&&+3a_sm_0^2\frac{x}{s}-12a_ss(1-x)(1-2x)+\frac32a_sm_0^2(4+x)\nonumber\\
&&-\frac{b}{4}m_s\frac{(1-x)(2+5x)}{x}\big\}e^{-\frac{s}{{M_B}^2}}ds,
\end{eqnarray}
and
\begin{eqnarray}
(2\pi)^4(\lambda_1\lambda_3^*+\lambda_1^*\lambda_3)M^2e^{-\frac{M^2}{{M_B}^2}}&=&\int_{4m_s^2}^{s_0}e^{-\frac{s}{{M_B}^2}}ds\big\{-\frac{m_s}{2}s^2[(1-x)(3+15x+4x^2)\nonumber\\
&&+12x(1+x)\ln{x}]-\frac12a_ss(1-x)(8-5x-5x^2)\nonumber\\
&&+\frac{1}{24}a_sm_0^2\frac{1}{1-x}[2x(2+x+x^2)+3(2-x^2+3x^3)]\nonumber\\
&&+\frac{1}{64}m_sb[-9(2-x)(1-x)+7\ln{x}]\nonumber\\
&&-\frac1{24}m_sb\frac{1}{x}[(1-x)(8+11x+11x^2)-18x\ln{x}]\big\}.
\end{eqnarray}

The numerical analysis shows that if $f_\Xi$ is taken to be positive, the final numerical values of the coupling constants of $\Xi$ are given as follows:
\begin{eqnarray}
f_{\Xi}&=&(9.9\pm0.4)\times10^{-3}\; \mbox{GeV}^2,\hspace{2.5cm}\lambda_1=-(2.8\pm0.1)\times10^{-2}\; \mbox{GeV}^2,\nonumber\\
\lambda_2&=&(5.2\pm0.2)\times10^{-2}\; \mbox{GeV}^2,\hspace{2.5cm}\lambda_3=(1.7\pm0.1)\times10^{-2}\; \mbox{GeV}^2.\label{sigmapara}
\end{eqnarray}

\section{Application: form factors of the baryons with light-cone QCD sum rules}\label{sec:app}
\subsection{Electromagnetic form factors}\label{EMff}
\subsubsection{LCSRs for the EM form factors}
Electromagnetic form factors, which characterize the internal structure of the hadron, have received great attention in the past decades. There have been a lot of
investigations both experimentally and theoretically for mesons. However, as three-body composite particles, baryons have more complex structure in comparison with
mesons. Thus studies on the baryon EM form factors are much fewer than those of the meson's. The existing investigations are mainly focused on the nucleon
\cite{Lenz,Aliev} (recent status can be found in Ref. \cite{overviewofDAs} and references therein). We have given an investigation on the EM form factors of the
$\Lambda$ and $\Sigma$ baryons in previous work \cite{DAs,LYL,Ioffeff}. In this subsection, the EM form factors of the $\Xi$ baryon are studied by use of the
light-cone QCD sum rule method; furthermore, the magnetic moments of the same baryons are estimated by fitting our results with the dipole formula.

The definition of the EM form factors is connected with the matrix element of the EM current of a baryon between the baryon states:
\begin{equation}
\langle\Xi(P,s)|j_\mu^{em}(0)|\Xi(P',s')\rangle=\bar \Xi (P,s)[\gamma_\mu F_1(Q^2)-i\frac{\sigma_{\mu\nu} q^\nu}{2M}F_2(Q^2)]\Xi(P',s'),\label{form}
\end{equation}
where $Q^2=-q^2=-(P-P')^2$ is the squared momentum transfer, $F_1(Q^2)$ and $F_2(Q^2)$ are the Dirac and Pauli form factors, respectively, and $j_\mu^{em}=e_u\bar
u\gamma_\mu u+e_s \bar s\gamma_\mu s$ is the electromagnetic current relevant to the hadron. $P,s$ and $P',s'$ are the four-momenta and the spins of the initial and
the final $\Xi$ baryon states, respectively. Experimentally, the EM form factors are usually expressed by the electric $G_E(Q^2)$ and magnetic $G_M(Q^2)$ Sachs form
factors:
\begin{eqnarray}
G_E(Q^2)&=&F_1(Q^2)-\frac{Q^2}{4M^2}F_2(Q^2),\nonumber\\
G_M(Q^2)&=&F_1(Q^2)+F_2(Q^2).
\end{eqnarray}
The normalization of the magnetic form factor $G_M(Q^2)$ at the point $Q^2=0$ gives the magnetic moment of the baryon:
\begin{equation}
G_M(0)=\mu_\Xi.
\end{equation}

In the following we mainly focus on the $\Xi^0$ baryon, and the calculation of $\Xi^-$ is similar. The derivation begins with the following correlation function:
\begin{equation}
T_\mu(P,q)=i \int d^4xe^{iq\cdot x}\langle 0|T\{j_{\Xi^0}(0)j_\mu^{em}(x)\}|\Xi^0(P,s)\rangle,\label{correlator}
\end{equation}
where the interpolating current is used as the form (\ref{CZcurrent}). The hadronic representation of the correlation function is acquired by inserting a complete
set of states with the same quantum numbers as those of $\Xi^0$:
\begin{eqnarray}
z^\mu T_\mu(P,q)&=&\frac{1}{M_{\Xi^0}^2-P'^2}f_{\Xi^0} (P'\cdot z)[2(P'\cdot zF_1(Q^2)-\frac{q\cdot z}{2}F_2(Q^2))\!\not\! z\nonumber\\&&+(P'\cdot
zF_2(Q^2)+\frac{q\cdot z}{2}F_2(Q^2))\frac{\!\not\! z\!\not\! q}{M_{\Xi^0}}]\Xi^0(P,s)+...,
\end{eqnarray}
in which $P'=P-q$, and the dots stand for the higher resonances and continuum contributions. Herein the correlation function is contracted with the light-cone
vector $z^\mu$ to get rid of contributions proportional to $z^\mu$ which are subdominant on the light cone. On the theoretical side, the correlation function
(\ref{correlator}) can be calculated with the aid of the DAs obtained above to the leading order of $\alpha_s$:
\begin{eqnarray}
z_\mu T^\mu &=&2(P\cdot z)^2(\!\not\! {z}\Xi)_\gamma \Big\{e_u\int_0^1d\alpha_3\frac{1}{s-p'^2}\{B_0(\alpha_3)+\frac{M^2}{(s-p'^2)}B_1(\alpha_3)
\nonumber\\
&&-2\frac{M^4}{(s-P'^2)^2}B_2(\alpha_3)\}+2e_s\int_0^1d\alpha_2\frac{1}{s_2-P'^2}\{C_0(\alpha_2)
\nonumber\\
&&+\frac{M^2}{(s_2-P'^2)}C_1(\alpha_2)-2\frac{M^4}{(s_2-P'^2)^2}C_2(\alpha_2)\} \Big\}\nonumber\\
&&+2(P\cdot
z)^2M(\!\not\!{z}\!\not\!{q}\Xi)_\gamma\Big\{e_u\int_0^1d\alpha_3\frac{1}{\alpha_3(s-P'^2)^2}\{-D_1(\alpha_3)\nonumber\\
&&+2\frac{M^2}{(s-P'^2)}B_2(\alpha_3)\}+2e_s\int_0^1d\alpha_2\frac{1}{\alpha_2(s_2-P'^2)^2}\{-E_1(\alpha_2)\nonumber\\
&& +2\frac{M^2}{(s_2-P'^2)}C_2(\alpha_2)\}\Big\},
\end{eqnarray}
where $s=(1-\alpha_3)M^2+\frac{(1-\alpha_3)}{\alpha_3}Q^2$, $s_2=(1-\alpha_2)M^2+\frac{(1-\alpha_2)}{\alpha_2}Q^2+\frac{m_s^2}{\alpha_2}$, and
\begin{eqnarray}
B_0(\alpha_3)&=&\int_0^{1-\alpha_3}d\alpha_1V_1(\alpha_1,1-\alpha_1-\alpha_3,\alpha_3),\nonumber\\
B_1(\alpha_3)&=&(2\widetilde V_1-\widetilde V_2-\widetilde V_3-\widetilde V_4-\widetilde V_5)(\alpha_3),\nonumber\\
B_2(\alpha_3)&=&(-\widetilde{\widetilde V_1}+\widetilde{\widetilde V_2}+\widetilde{\widetilde V_3}+\widetilde{\widetilde V_4}+\widetilde{\widetilde
V_5}-\widetilde{\widetilde V_6})(\alpha_3),\nonumber\\
C_0(\alpha_2)&=&\int_0^{1-\alpha_2}d\alpha_1V_1(\alpha_1,\alpha_2,1-\alpha_1-\alpha_2),\nonumber\\
C_1(\alpha_2)&=&(2\widetilde V_1-\widetilde V_2-\widetilde V_3-\widetilde V_4-\widetilde V_5)(\alpha_2),\nonumber\\
C_2(\alpha_2)&=&(-\widetilde{\widetilde V_1}+\widetilde{\widetilde V_2}+\widetilde{\widetilde V_3}+\widetilde{\widetilde V_4}+\widetilde{\widetilde
V_5}-\widetilde{\widetilde V_6})(\alpha_2),\nonumber\\
D_1(\alpha_3)&=&(\widetilde V_1-\widetilde V_2-\widetilde V_3)(\alpha_3),\nonumber\\
E_1(\alpha_2)&=&(\widetilde V_1-\widetilde V_2-\widetilde V_3)(\alpha_2),
\end{eqnarray}
in which
\begin{eqnarray}
\widetilde V_i(\alpha_2)&=&\int_0^{\alpha_2}d{\alpha_2'}\int_0^{1-\alpha_2'}d\alpha_1V_i(\alpha_1,\alpha_2',1-\alpha_1-\alpha_2'),\nonumber\\
\widetilde{\widetilde V_i}(\alpha_2)&=&\int_0^{\alpha_2}d{\alpha_2'}\int_0^{\alpha_2'}d{\alpha_2''}\int_0^{1-\alpha_2''}d\alpha_1V_i(\alpha_1,\alpha_2'',1-\alpha_1-\alpha_2''),\nonumber\\
\widetilde V_i(\alpha_3)&=&\int_0^{\alpha_3}d{\alpha_3'}\int_0^{1-\alpha_3'}d\alpha_1V_i(\alpha_1,1-\alpha_1-\alpha_3',\alpha_3'),\nonumber\\
\widetilde{\widetilde
V_i}(\alpha_3)&=&\int_0^{\alpha_3}d{\alpha_3'}\int_0^{\alpha_3'}d{\alpha_3''}\int_0^{1-\alpha_3''}d\alpha_1V_i(\alpha_1,1-\alpha_1-\alpha_3'',\alpha_3'').
\end{eqnarray}
By matching both sides of the Borel transformed version of the hadronic and theoretical representations with the assumption of quark-hadron duality, the final sum
rules are given as follows,
\begin{eqnarray}
f_{\Xi^0}F_1(Q^2)e^{-\frac{M^2}{{M_B}^2}}&=&e_u\int_{\alpha_{30}}^1d\alpha_3e^{-\frac{s}{{M_B}^2}}\Big\{B_0(\alpha_3)
+\frac{M^2}{{M_B}^2}B_1(\alpha_3)-\frac{M^4}{{M_B}^4}B_2(\alpha_3)\Big\}\nonumber\\
&&+e_ue^{-\frac{s_0}{{M_B}^2}}\frac{\alpha_{30}^2M^2}{\alpha_{20}^2M^2+Q^2}\Big\{B_1(\alpha_{30})-\frac{M^2}{M_B^2}B_2(\alpha_{30})\Big\}\nonumber\\
&&+e_ue^{-\frac{s_0}{{M_B}^2}}\frac{\alpha_{30}^2M^4}{\alpha_{30}^2M^2+Q^2}\frac{d}{d\alpha_{30}}
B_2(\alpha_{30})\frac{\alpha_{30}^2}{\alpha_{30}^2M^2+Q^2}\nonumber\\
&&+2e_s\int_{\alpha_{20}}^1d\alpha_2e^{-\frac{s_2}{{M_B}^2}}\Big\{C_0(\alpha_2)
+\frac{M^2}{{M_B}^2}C_1(\alpha_2)-\frac{M^4}{{M_B}^4}C_2(\alpha_2)\Big\}\nonumber\\
&&+2e_se^{-\frac{s_0}{{M_B}^2}}\frac{\alpha_{20}^2M^2}{\alpha_{20}^2M^2+Q^2+m_s^2}\Big\{C_1(\alpha_{20})-\frac{M^2}{M_B^2}C_2(\alpha_{20})\Big\}\nonumber\\
&&+2e_se^{-\frac{s_0}{{M_B}^2}}\frac{\alpha_{20}^2M^4}{\alpha_{20}^2M^2+Q^2+m_s^2}\frac{d}{d\alpha_{20}}
C_2(\alpha_{20})\frac{\alpha_{20}^2}{\alpha_{20}^2M^2+Q^2+m_s^2},\nonumber\\
\end{eqnarray}
\begin{eqnarray}
f_{\Xi^0}F_2(Q^2)e^{-\frac{M^2}{{M_B}^2}}&=&M^2\Big\{2e_u\int_{\alpha_{30}}^1d\alpha_3e^{-\frac{s}{{M_B^2}}}\frac{1}{\alpha_3{M_B}^2}\big\{-D_1(\alpha_3)
+\frac{M^2}{{M_B}^2}B_2(\alpha_3)\big\}
\nonumber\\&&-2e_ue^{-\frac{s_0}{{M_B}^2}}\frac{\alpha_{30}}{\alpha_{30}^2M^2+Q^2}\big\{D_1(\alpha_{30})-\frac{M^2}{M_B^2}B_2(\alpha_{30})\big\}\nonumber\\
&&-2e_ue^{-\frac{s_0}{{M_B}^2}}\frac{\alpha_{30}^2M^2}{\alpha_{30}^2M^2+Q^2}
\frac{d}{d\alpha_{30}}B_2(\alpha_{30})\frac{\alpha_{30}}{\alpha_{30}^2M^2+Q^2}\nonumber\\
&&+4e_s\int_{\alpha_{20}}^1d\alpha_{2}e^{-\frac{s_2}{{M_B}^2}}\frac{1}{\alpha_2{M_B}^2}\big\{-E_1(\alpha_2)+\frac{M^2}{{M_B}^2}C_2(\alpha_2)\big\}\nonumber\\
&&-4e_se^{-\frac{s_{0}}{M_B^2}}\frac{\alpha_{20}}{\alpha_{20}^2M^2+Q^2+m_s^2}\big\{E_1(\alpha_{20})-\frac{M^2}{M_B^2}C_2(\alpha_{20})\big\}\nonumber\\
&&-4e_se^{-\frac{s_{0}}{M_B^2}}\frac{\alpha_{20}^2M^2}{\alpha_{20}^2M^2+Q^2+m_s^2}\frac{d}{d\alpha_{20}}C_2(\alpha_{20})
\frac{\alpha_{20}}{\alpha_{20}^2M^2+Q^2+m_s^2}\Big \},\nonumber\\
\end{eqnarray}
where $\alpha_{i0}$ corresponds to the continuum threshold $s_0$ by the following expressions:
\begin{eqnarray}
\alpha_{20}&=&\frac{-(Q^2+s_0-M^2)+\sqrt{(Q^2+s_0-M^2)^2+4(Q^2+m_s^2)M^2}}{2M^2},\nonumber\\
\alpha_{30}&=&\frac{-(Q^2+s_0-M^2)+\sqrt{(Q^2+s_0-M^2)^2+4Q^2M^2}}{2M^2}.\label{threshold}
\end{eqnarray}

\subsubsection{Numerical analysis}

\begin{figure}
\begin{minipage}{7cm}
\epsfxsize=6cm \centerline{\epsffile{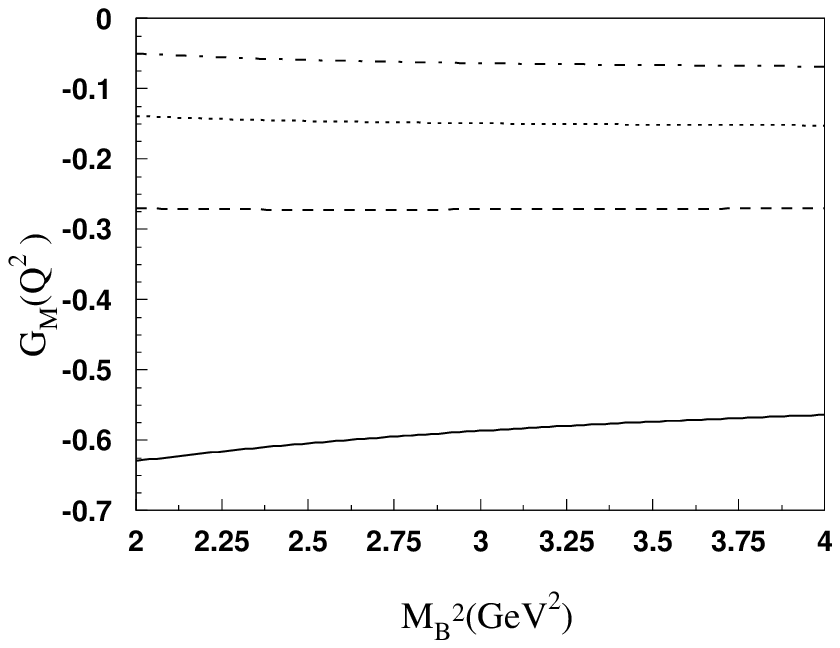}}
\end{minipage}
\hfill
\begin{minipage}{7cm}
\epsfxsize=6cm \centerline{\epsffile{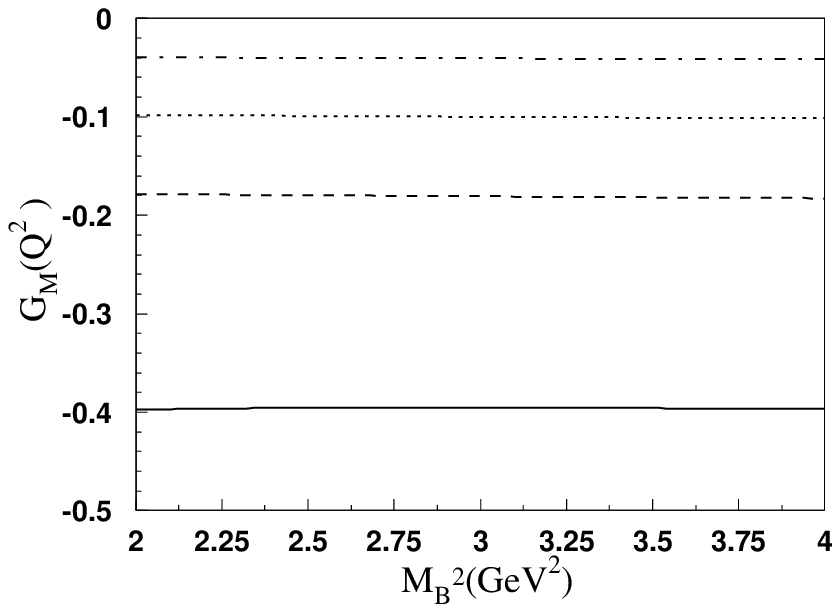}}
\end{minipage}
\caption{\quad Dependence of the magnetic form factor $G_M(Q^2)$ of $\Xi$ baryons on the Borel parameter at different momentum transfer. The lines correspond to the
points $Q^2=1\,,2\,,3\,,5\;\ \mbox{GeV}^{2}$ from the bottom up for $\Xi^0$ (left) and $\Xi^-$ (right), respectively.}\label{fq}
\end{figure}

In the numerical analysis, the continuum threshold is chosen to be $s_0=(2.8-3.0)\, \mbox{GeV}^2$; the masses of the $\Xi$ baryons are from Ref. \cite{PDG}:
$M_{\Xi^0}=1.315\;\ \mbox{GeV}$ and $M_{\Xi^-}=1.322\;\ \mbox{GeV}$. To choose the working window of the Borel parameter, we first give the dependence of the EM
form factors on the Borel parameter at different points of $Q^2$ in Fig. \ref{fq}. It can be seen from the panels that there is an acceptable "stability window" in
the range $2\; \mbox{GeV}^2\leq M_B^2 \leq 4\; \mbox{GeV}^2$. Hereafter, the Borel parameter is set to be $M_B^2=3\;\mbox{GeV}^2$ in the following analysis.

\begin{figure}
\begin{minipage}{7cm}
\epsfxsize=6cm \centerline{\epsffile{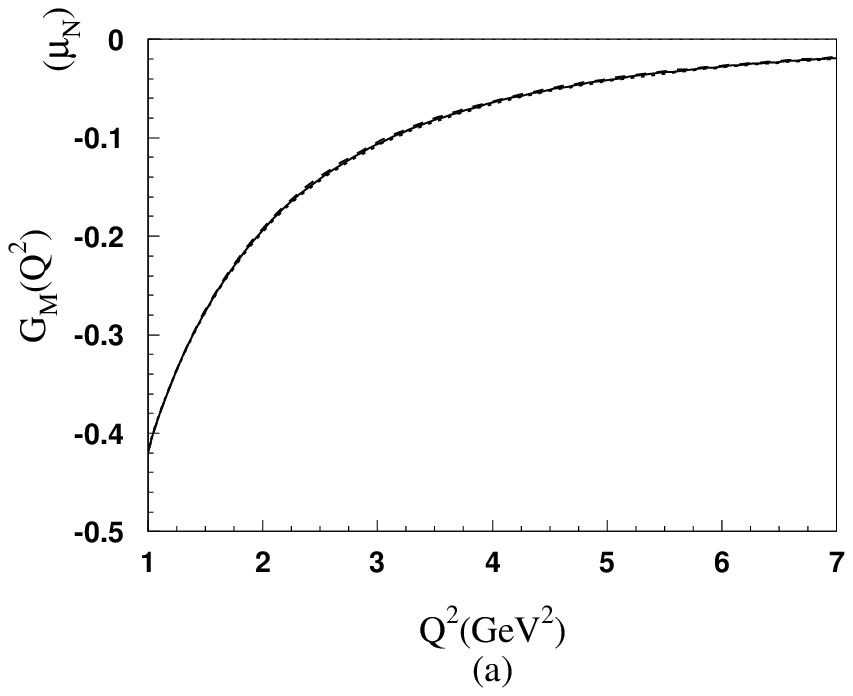}}
\end{minipage}
\hfill
\begin{minipage}{7cm}
\epsfxsize=6cm \centerline{\epsffile{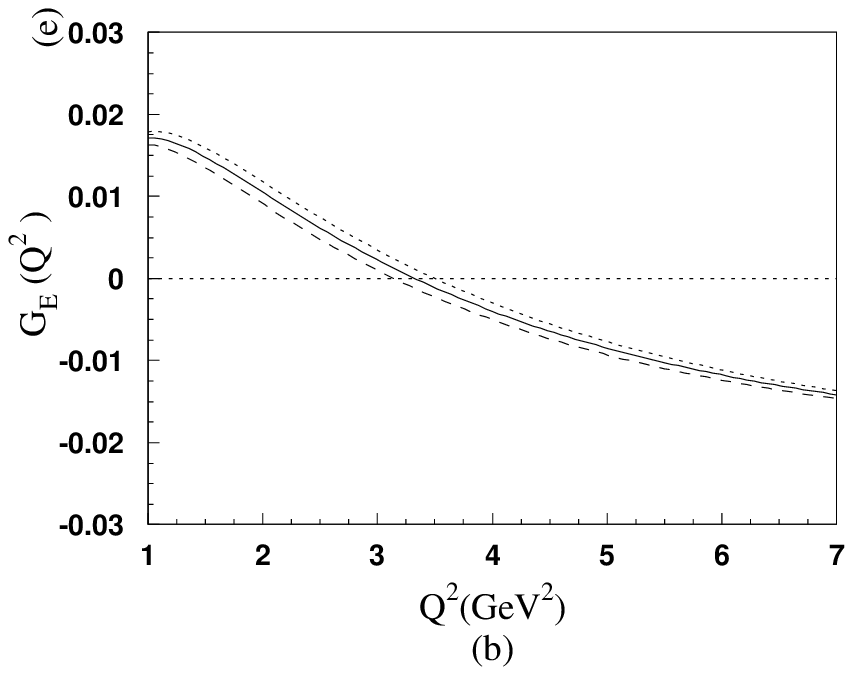}}
\end{minipage}
\caption{\quad $Q^2$ dependence of the magnetic form factor (a) and the electronic form factor (b) of $\Xi^0$. The lines correspond to the threshold
$s_0=2.8\,,2.9\,,3.0\; \mbox{GeV}^2$ from the bottom up (a) and from the top down (b).}\label{xi}
\end{figure}

\begin{figure}
\begin{minipage}{7cm}
\epsfxsize=6cm \centerline{\epsffile{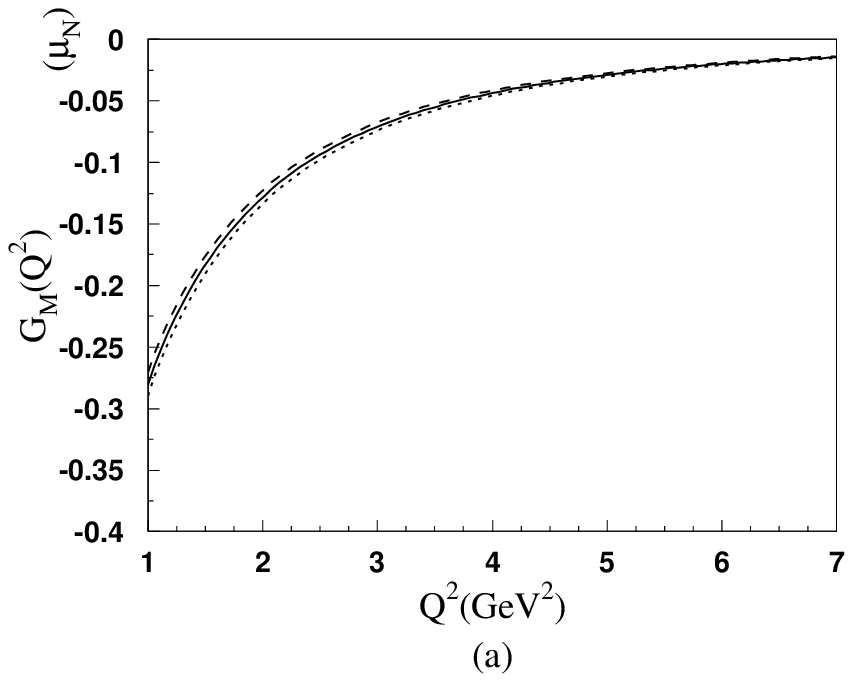}}
\end{minipage}
\hfill
\begin{minipage}{7cm}
\epsfxsize=6cm \centerline{\epsffile{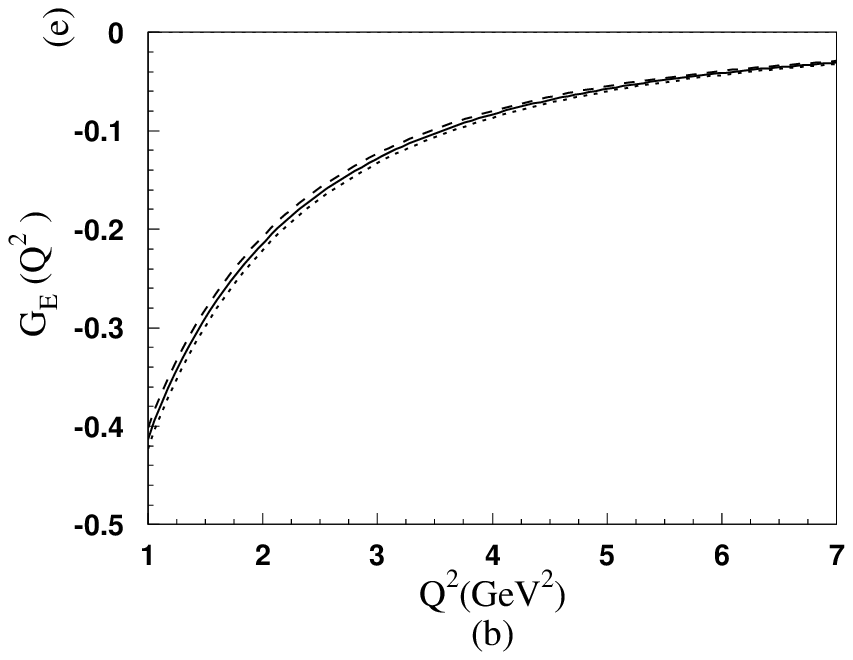}}
\end{minipage}
\caption{\quad $Q^2$ dependence of the magnetic form factor (a) and the electronic form factor (b) of $\Xi^-$. The lines correspond to the threshold
$s_0=2.8\,,2.9\,,3.0\; \mbox{GeV}^2$ from the top down.}\label{negxi}
\end{figure}

The $Q^2$-dependent magnetic and electric form factors are shown in Fig. \ref{xi} for $\Xi^0$ and Fig. \ref{negxi} for $\Xi^-$. It is known that LCSR needs the
momentum transfer $Q^2$ to be sufficiently large which is due to the convergence of the light-cone expansion. The calculation shows that higher twist contributions
are suppressed efficiently above the point $Q^2=1\,\mbox{GeV}^2$, and the higher resonance contributions are subdominant after Borel transformation below
$Q^2=7\,\mbox{GeV}^2$. Therefore we carry on the numerical analysis in the range $1\,\mbox{GeV}^2\le Q^2\le7\,\mbox{GeV}^2$. In comparison with results from the
constituent quark model in Ref. \cite{Cauteren}, our calculations are in accordance with the authors' but for the electronic form factor of $\Xi^0$. In our
calculation, the electric form factor turns from positive to negative when $Q^2$ increases while their conclusion is the opposite. Taking into account calculations
from the chiral quark/soliton model in Ref. \cite{Kim}, in which behaviors of the EM form factors below $Q^2=1\,\mbox{GeV}^2$ are given, our result is reasonable.

\begin{figure}
\begin{minipage}{7cm}
\epsfxsize=6cm \centerline{\epsffile{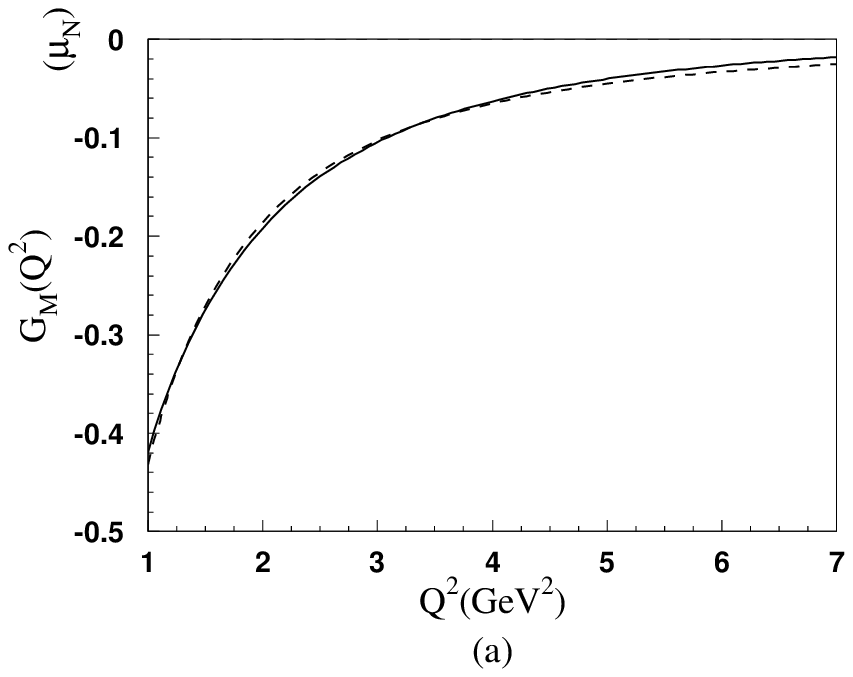}}
\end{minipage}
\hfill
\begin{minipage}{7cm}
\epsfxsize=6cm \centerline{\epsffile{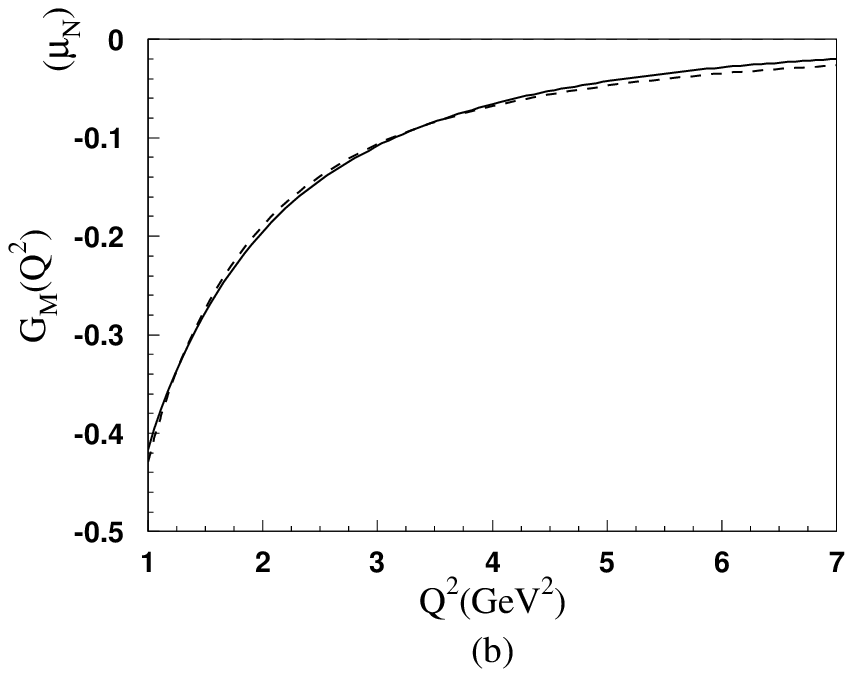}}
\end{minipage}
\caption{\quad Fits of the magnetic form factor $G_M(Q^2)$ of $\Xi^0$ by the dipole formula $\mu_{\Xi^0}/(1+Q^2/m_0^2)^2$. The dashed lines are the fits, and
$(a)\,,(b)$ correspond to the threshold $s_0=2.8\,,3.0\; \mbox{GeV}^2$, respectively.}\label{fit}
\end{figure}

\begin{figure}
\begin{minipage}{7cm}
\epsfxsize=6cm \centerline{\epsffile{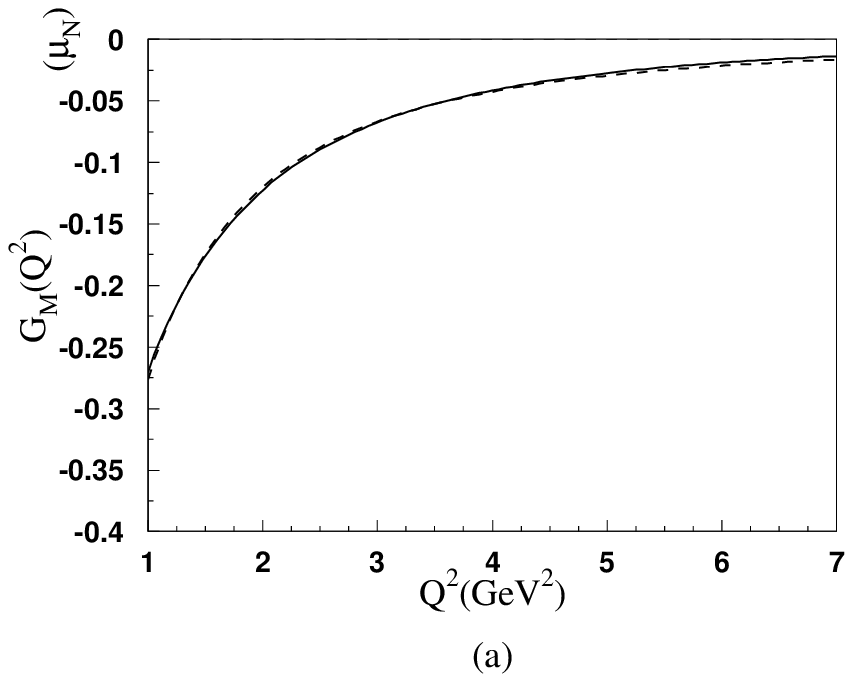}}
\end{minipage}
\hfill
\begin{minipage}{7cm}
\epsfxsize=6cm \centerline{\epsffile{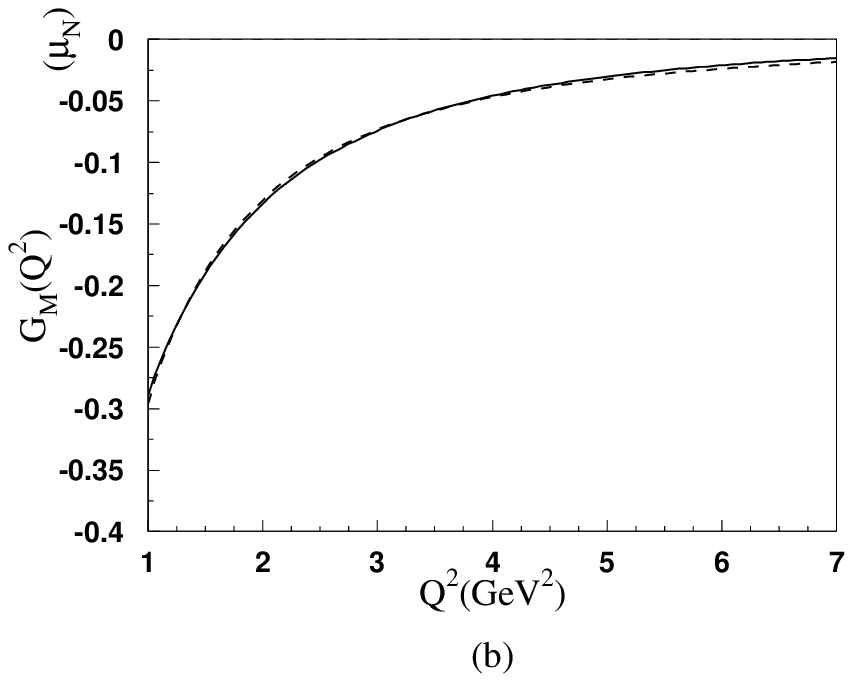}}
\end{minipage}
\caption{\quad Fits of the magnetic form factor $G_M(Q^2)$ of $\Xi^-$ by the dipole formula $\mu_{\Xi^-}/(1+Q^2/m_0^2)^2$. The dashed lines are the fits, and
$(a)\,,(b)$ correspond to the threshold $s_0=2.8\,,3.0\; \mbox{GeV}^2$, respectively.}\label{negfit}
\end{figure}

Similar to the cases of the nucleon and other octet baryons \cite{DAs,LYL,Lenz}, the magnetic form factors are assumed to be expressed by the dipole formula:
\begin{equation}
\frac{1}{\mu_\Xi}G_M(Q^2)=\frac{1}{(1+Q^2/m_0^2)^2}=G_D(Q^2).\label{dipole}
\end{equation}
As there is no information about the parameter $m_0^2$ from experimental data so far, the two parameters $m_0^2$ and $\mu_\Xi$ are estimated simultaneously by the
dipole formula (\ref{dipole}) fitting of the magnetic form factor. The simulation is shown by the dashed line in Fig. \ref{fit} for $\Xi^0$ and in Fig. \ref{negfit}
for $\Xi^-$. Our numerical values are $\mu_{\Xi^0}=-(1.92\pm0.34)\mu_N$ and $\mu_{\Xi^-}=-(1.19\pm0.03)\mu_N$. The uncertainties come from the different choice of
the threshold $s_0$ with the Borel parameter variation in the range $2.5\; \mbox{GeV}^2\leq M_B^2 \leq 3.5\; \mbox{GeV}^2$. The numerical analysis shows that the
calculation result for $\Xi^-$ is not sensitive to the threshold and the Borel parameter. Note that, as the estimations are from the dipole formula fits with the
sum rules as the input data, the uncertainty due to the variation of the input parameters, such as $f_{\Xi}$ and $\lambda_i$, is not included; it may reach 5\% or
more. In comparison with the values given by PDG \cite{PDG}, our estimations are larger in absolute values, which may partly lie in the accuracy of the DAs that are
calculated only to leading order conformal spin expansion. Another possibility is that in the dipole formula the uncertainty of the parameter $m_0^2$ gives a
deviation from the formula.

\begin{figure}
\begin{minipage}{7cm}
\epsfxsize=6cm \centerline{\epsffile{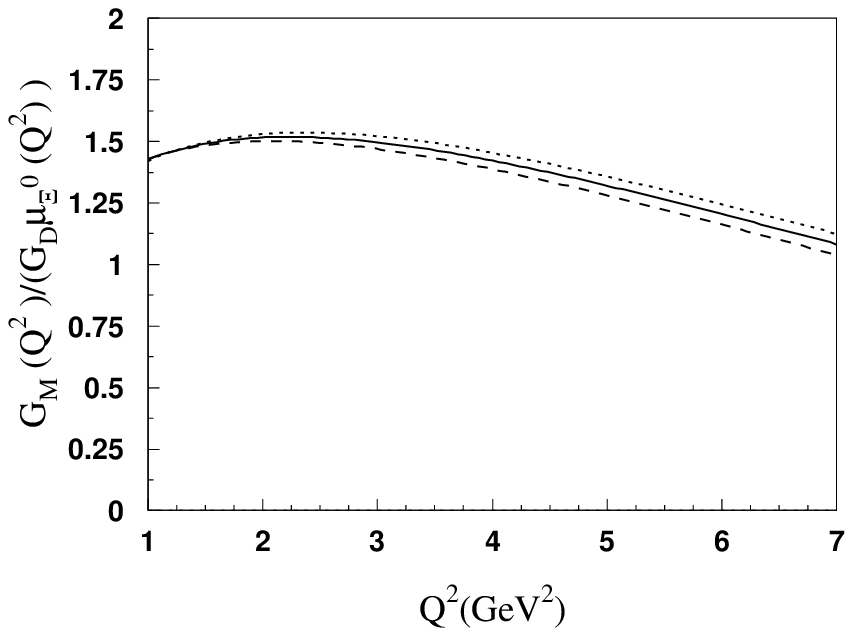}}
\end{minipage}
\hfill
\begin{minipage}{7cm}
\epsfxsize=6cm \centerline{\epsffile{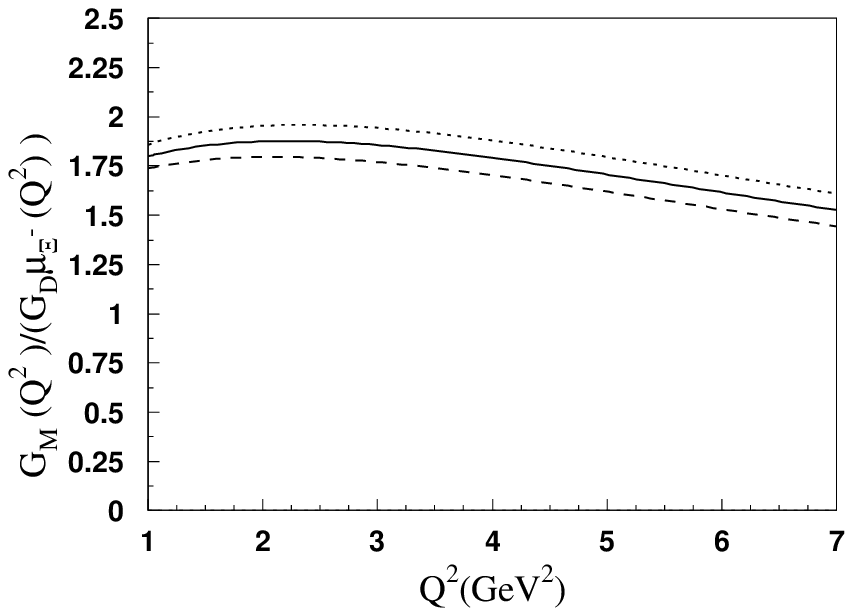}}
\end{minipage}
\caption{\quad The $Q^2$ dependence of the physical value $G_M/(\mu_\Xi G_D)$. The lines correspond to the threshold $s_0=2.8\,,2.9\,,3.0\; \mbox{GeV}^{2}$ from the
bottom up (left) and from the top down (right). The left corresponds to $\Xi^0$ and the right corresponds to $\Xi^-$.}\label{gmd}
\end{figure}

Finally, we plot the physical value $ G_M/(\mu_\Xi G_D)$ versus $Q^2$ in Fig. \ref{gmd}. The parameters in the dipole formula are chosen as follows: the magnetic
moments are used as the central value provided by PDG, $\mu_{\Xi^0}=-1.25\mu_N$ and $\mu_{\Xi^-}=-0.65\mu_N$; the other parameter $m_0^2$ is the central value from
the fits, $m_0^2=0.94$ for $\Xi^0$ and $m_0^2=0.96$ for $\Xi^-$. Future experiments are expected to provide more information about the EM form factors.

\subsection{Semileptonic decay $\Xi_c\rightarrow\Xi e\nu_e$}\label{semidecay}
\subsubsection{LCSRs for weak transition form factors}
The semileptonic decays of charmed hadrons are important decay modes in heavy flavor physics since they can give useful information on the Cabibbo-Kobiyash-Maskawa
matrix elements. However, because of the strong interaction in the hadronic bound state, the calculation of the processes needs some nonperturbative theoretical
approaches. Hereafter, we make use of LCSR technique to investigate the exclusive semileptonic decays of $\Xi_c$ involving $\Xi$ baryons.

One of the main decay channels for the baryon $\Xi_c$ is the weak decay $\Xi_c\rightarrow \Xi e^+\nu_e$, which was observed experimentally early in the 1990s
\cite{Alexander}. Theoretically, the process has been studied with various models \cite{Perez,Singleton,Hussain,Chy,Ivanov}. In the theoretical calculations, the
hadronic part of the process is generally parametrized by the nonperturbative form factors which are defined by the matrix element of the weak current between the
baryon states:
\begin{eqnarray}
\langle
\Xi_c(P')|j_\mu^w|\Xi(P)\rangle=\bar{\Xi_c}(P')[f_1\gamma_\mu-i\frac{f_2}{M}\sigma_{\mu\nu}q^\nu-(g_1\gamma_\mu+i\frac{g_2}{M}\sigma_{\mu\nu}q^\nu)\gamma_5]\Xi(P),
\label{weakff}
\end{eqnarray}
in which the mass of the positron is neglected. The differential decay rate is given as
\begin{eqnarray}
\frac{d\Gamma}{dq^2}&=&\frac{G_F^2|V_{cs}|^2}{192 \pi^3M_{\Xi_c}^5}q^2\sqrt{q_+^2q_-^2}\Big\{-6f_1f_2M_{\Xi_c}m_{+}{q_-}^2+6g_1g_2M_{\Xi_c}m_{-}q_+^2\nonumber\\
&&+f_1^2M_{\Xi_c}^2(\frac{m_+^2m_-^2}{q^2}+m_-^2-2(q^2+2M_{\Xi_c}M_\Xi))\nonumber\\
&&+g_1^2M_{\Xi_c}^2(\frac{m_+^2m_-^2}{q^2}+m_+^2-2(q^2-2M_{\Xi_c}M_\Xi))\nonumber\\
&&-f_2^2[-2m_+^2m_-^2+m_+^2q^2+q^2(q^2+4M_{\Xi_c}M_\Xi)]\nonumber\\
&&-g_2^2[-2m_+^2m_-^2+m_-^2q^2+q^2(q^2-4M_{\Xi_c}M_\Xi)]\Big\}, \label{decayrate}
\end{eqnarray}
where $m_{\pm}=M_{\Xi_c}\pm M_\Xi$ and $q_{\pm}^2=q^2-m_{\pm}^2$ are used for convenience.

The form factors in Eq. (\ref{weakff}) can be estimated in the framework of LCSR. The calculation begins with the following correlation function:
\begin{equation}
T_\mu(P,q)=i \int d^4xe^{iq\cdot x}\langle 0|T\{j_{\Xi_c}(0)j_\mu^{w}(x)\}|\Sigma^+(P,s)\rangle,\label{weakcorrelator}
\end{equation}
where the interpolating current for the $\Xi_c$ baryon is chosen as
\begin{equation}
j_{\Xi_c}(x)=\epsilon^{ijk}[s^i(x)C\!\not\! {z}c^j(x)]\gamma_5\!\not\! {z}q^k(x),
\end{equation}
and the weak current is
\begin{equation}
j_\mu^w(x)=\bar c(x)\gamma_\mu(1-\gamma_5)s(x).
\end{equation}

Following the standard philosophy of the QCD sum rule, the hadronic representation of the correlation function is written as
\begin{eqnarray}
z_\mu T^\mu&=&\frac{2f_{\Xi_c} (P'\cdot z)^2}{M_{\Xi_c}^2-P'^2}[f_1(q^2)\!\not\! z+\frac{1}{M}f_2(q^2)\!\not\! z\!\not\! q-\nonumber\\
&&(g_1(q^2)\!\not\!
z\gamma_5-\frac{1}{M}g_2(q^2)\!\not\! z\!\not\! q\gamma_5)]\Xi(P,s)+...\,.
\end{eqnarray}
On the theoretical side, the correlation function is expanded on the light cone and can be expressed by the DAs. After the standard process of LCSR, we get the
following results:
\begin{eqnarray}
f_{\Xi_c}f_1(q^2)e^{-\frac{M^2}{M_B^2}}&=&\int_{\alpha_{20}}^1d\alpha_2e^{-\frac{s_2}{M_B^2}}\big\{C_0(\alpha_2)+\frac{M^2}{M_B^2}C_1(\alpha_2)-\frac{M^4}{M_B^4}
C_2(\alpha_2)\big\}\nonumber\\
&&+e^{-\frac{s_0}{M_B^2}}\frac{{\alpha_{20}}^2M^2}{M^2{\alpha_{20}}^2-q^2+m_s^2}\big\{C_1(\alpha_{20})-\frac{M^2}{M_B^2}C_2(\alpha_{20})\big\}\nonumber\\
&&+e^{-\frac{s_0}{M_B^2}}\frac{{\alpha_{20}}^2M^4}{M^2{\alpha_{20}}^2-q^2+m_s^2}\frac{d}{d\alpha_{20}}\frac{\alpha_{20}^2}{\alpha_{20}^2M^2-q^2+m_s^2}C_2(\alpha_{20}),\label{wff1}
\end{eqnarray}
\begin{eqnarray}
f_{\Xi_c}f_2(q^2)e^{-\frac{M^2}{M_B^2}}&=&-\int_{\alpha_{20}}^1d\alpha_2e^{-\frac{s_2}{M_B^2}}\frac{MM_{\Xi_c}}{M_B^2}\frac{1}{\alpha_2}\big\{E_1(\alpha_2)
-\frac{M^2}{M_B^2}C_2(\alpha_2)\big\}\nonumber\\
&&-e^{-\frac{s_0}{M_B^2}}\frac{{\alpha_{20}}MM_{\Xi_c}}{M^2{\alpha_{20}}^2-q^2+m_s^2}\big\{E_1(\alpha_{20})-\frac{M^2}{M_B^2}C_2(\alpha_{20})\big\}\nonumber\\
&&-e^{-\frac{s_0}{M_B^2}}\frac{{\alpha_{20}}^2M^3M_{\Xi_c}}{M^2{\alpha_{20}}^2-q^2+m_s^2}\frac{d}{d\alpha_{20}}\frac{\alpha_{20}}{\alpha_{20}^2M^2-q^2+m_s^2}C_2(\alpha_{20}),\label{wff2}
\end{eqnarray}
\begin{eqnarray}
f_{\Xi_c}g_1(q^2)e^{-\frac{M^2}{M_B^2}}&=&\int_{\alpha_{20}}^1d\alpha_2e^{-\frac{s_2}{M_B^2}}\big\{H_1(\alpha_2)+\frac{M^2}{M_B^2}H_2(\alpha_2)+\frac{M^4}{M_B^4}H_3(\alpha_2)\big\}\nonumber\\
&&+e^{-\frac{s_0}{M_B^2}}\frac{{\alpha_{20}}^2M^2}{M^2{\alpha_{20}}^2-q^2+m_s^2}\big\{H_2(\alpha_{20})+\frac{M^2}{M_B^2}H_3(\alpha_{20})\big\}\nonumber\\
&&-e^{-\frac{s_0}{M_B^2}}\frac{{\alpha_{20}}^2M^4}{M^2{\alpha_{20}}^2-q^2+m_s^2}\frac{d}{d\alpha_{20}}\frac{\alpha_{20}^2}{\alpha_{20}^2M^2-q^2+m_s^2}H_3(\alpha_{20}),\label{wff3}
\end{eqnarray}

\begin{eqnarray}
f_{\Xi_c}g_2(q^2)e^{-\frac{M^2}{M_B^2}}&=&\int_{\alpha_{20}}^1d\alpha_2e^{-\frac{s_2}{M_B^2}}\frac{MM_{\Xi_c}}{M_B^2}\frac{1}{\alpha_2}\big\{H_4(\alpha_2)
+\frac{M^2}{M_B^2}H_3(\alpha_2)\big\}\nonumber\\
&&+e^{-\frac{s_0}{M_B^2}}\frac{{\alpha_{20}}MM_{\Xi_c}}{M^2{\alpha_{20}}^2-q^2+m_s^2}\big\{H_4(\alpha_{20})+\frac{M^2}{M_B^2}H_3(\alpha_{20})\big\}\nonumber\\
&&-e^{-\frac{s_0}{M_B^2}}\frac{{\alpha_{20}}^2M^3M_{\Xi_c}}{M^2{\alpha_{20}}^2-q^2+m_s^2}\frac{d}{d\alpha_{20}}\frac{\alpha_{20}}{\alpha_{20}^2M^2-q^2+m_s^2}H_3(\alpha_{20}),\label{wff4}
\end{eqnarray}
in which the following additional notations are used for convenience,
\begin{eqnarray}
H_1(\alpha_2)&=&\int_0^{\alpha_2}d\alpha_1A_1(\alpha_1,\alpha_2,1-\alpha_1-\alpha_2),\nonumber\\
H_2(\alpha_2)&=&(\widetilde A_2-\widetilde A_3+\widetilde A_4-\widetilde A_5)(\alpha_2),\nonumber\\
H_3(\alpha_2)&=&(\widetilde{\widetilde A_1}-\widetilde{\widetilde A_2}+\widetilde{\widetilde A_3}+\widetilde{\widetilde A_4}-\widetilde{\widetilde
A_5}+\widetilde{\widetilde A_6})(\alpha_2),\nonumber\\
H_4(\alpha_2)&=&(-\widetilde A_1+\widetilde A_2-\widetilde A_3)(\alpha_2).
\end{eqnarray}
The threshold-related parameter $\alpha_{20}$ is the same as that in Eqs. (\ref{threshold}) with the interchange $m_s\leftrightarrow m_c$.
\subsubsection{Numerical analysis}

Before the numerical analysis, we first specify the choice of the parameters used in this subsection. The continuum threshold is chosen as $s_0=(7-9)\,
\mbox{GeV}^2$, and the masses of the baryons can be found in Ref. \cite{PDG}. As in the sum rules, the two $\Xi_c$ baryons belong to the same isomultiplet, effects
from the isospin symmetry breaking can be neglected safely. Thus we take the mass of $\Xi_c$ as $M_{\Xi_c}=2.471\;\mbox{GeV}$.

As to the coupling constant $f_{\Xi_c}$, we use the QCD sum rule method to estimate it. Complying with the standard procedure of the QCD sum rule, we get the
following expression:
\begin{eqnarray}
(4\pi)^4f_{\Xi_c}^2&=&\int_{(m_s+m_b)^2}^{s_0}dse^{-\frac{s-M_{\Xi_c}^2}{M_B^2}}\big\{\frac25s(1-x)^5-8m_sa_s\frac1sx^2(1-x)\nonumber\\
&&+\frac{1}{12}m_sa_sm_0^2\frac{1}{s^2}x(2-3x)-\frac16b\frac{1}{s}x(1-x)^2\big\},
\end{eqnarray}
where $x=m_c/s$ is used for convenience. In the numerical analysis, the threshold is $7\; \mbox{GeV}^2\leq s_0 \leq 9\; \mbox{GeV}^2$ and the Borel window is $1.5\;
\mbox{GeV}^2\leq M_B^2 \leq 2\; \mbox{GeV}^2$. The $c$-quark mass is taken as the central value from PDG: $m_c=1.27\; \mbox{GeV}$. Other parameters are used the
same ones given in Sec. \ref{sec:sumrule}. Our estimation for the coupling constant is $f_{\Xi_c}=(8.6\pm 0.9)\times10^{-3}\,\mbox{GeV}^2$.

The Borel parameter is chosen to suppress both higher resonance and higher twist contributions. The calculations show that the results are acceptable in the range
$7\; \mbox{GeV}^2\leq M_B^2 \leq 9\; \mbox{GeV}^2$. In the following analysis, the Borel parameter is set to be $M_B^2=8\;\mbox{GeV}^2$.

\begin{figure}
\begin{minipage}{7cm}
\epsfxsize=6cm \centerline{\epsffile{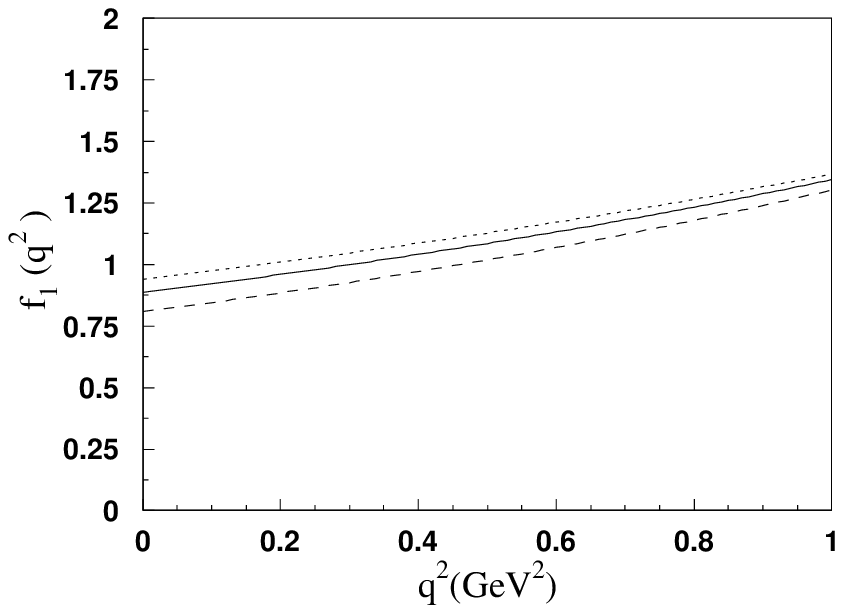}}
\end{minipage}
\hfill
\begin{minipage}{7cm}
\epsfxsize=6cm \centerline{\epsffile{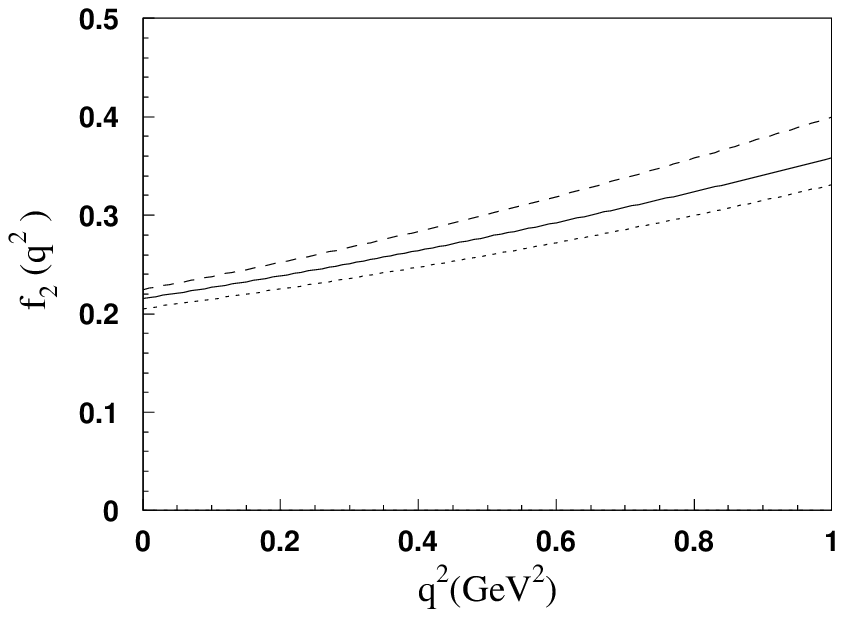}}
\end{minipage}
\hfill
\begin{minipage}{7cm}
\epsfxsize=6cm \centerline{\epsffile{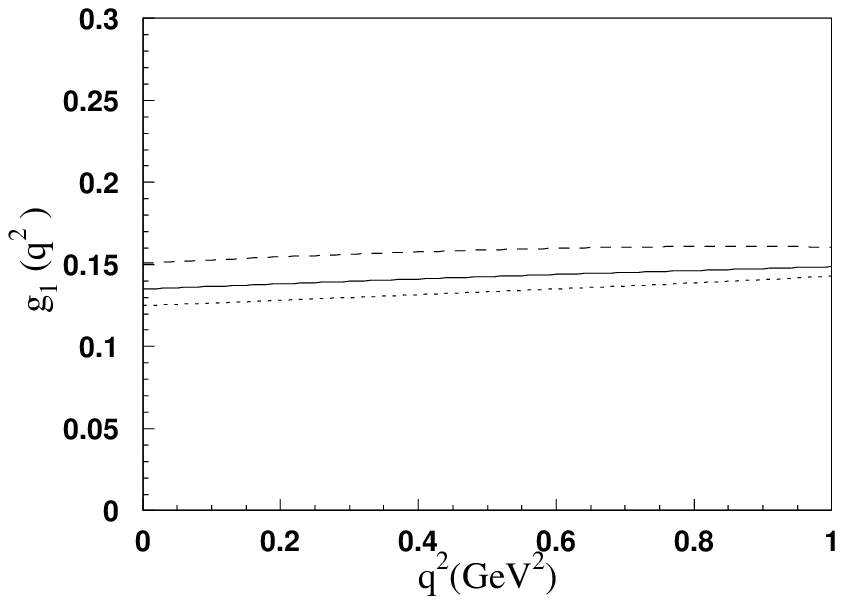}}
\end{minipage}\hfill
\begin{minipage}{7cm}
\epsfxsize=6cm \centerline{\epsffile{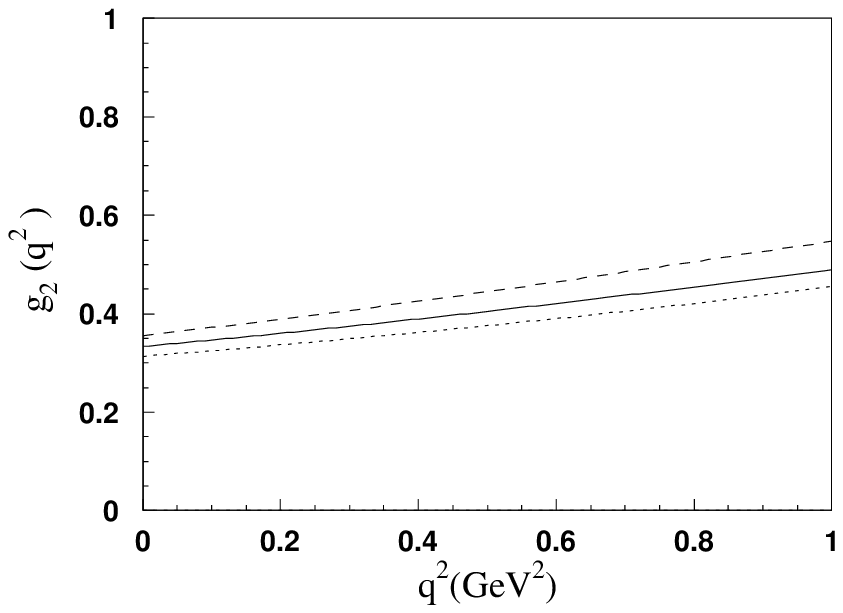}}
\end{minipage}\caption{\quad $q^2$ dependence of the weak form factors.
The lines correspond to the threshold $s_0=7\,, 8\,,9\; \mbox{GeV}^2$ from the top down for $f_i(q^2)$ and from the bottom up for $g_i(q^2)$.}\label{weakf}
\end{figure}

We plot the weak form factors depending on the momentum transfer $q^2$ in Fig. \ref{weakf}. Note that unlike the case for EM form factors in which the momentum
transfer is spacelike, in the decay the physical process can only occur in the timelike region. At the same time, LCSR needs the momentum transfer to satisfy the
relation $q^2-m_c^2\ll0$ so that the main contribution is dominated on the light cone. Hence we choose the range of $q^2$ as $0\leq q^2\leq 1.0\,\mbox{GeV}^2$ in
the sum rules.

The thorough investigation of the decay process needs to determine behaviors of the form factors in the whole physical region, $0\leq q^2\leq (M_{\Xi_c}-M_\Xi)^2$.
To this end, we recast the form factors as explicit functions of timelike momentum transfer by the dipole formula fits in the sum rule allowed range $0\leq q^2\leq
1.0\,\mbox{GeV}^2$, and then extrapolate them to the whole kinematical region. The three-parameter dipole formula is written as
\begin{equation}
f(q^2)=\frac{f(0)}{1+a(q^2/{M_{\Xi_c}}^2)+b(q^2/{M_{\Xi_c}}^2)^2},\label{weakformulafit}
\end{equation}
where $f(q^2)$ represents the form factors $f_i(q^2)$ or $g_i(q^2)$ ($i=1,2$). The coefficients are listed in Table \ref{weakfit}.
\begin{table}
\caption{Fits of the weak form factors by the dipole formula (\ref{weakformulafit}).}
\begin{center}
\begin{tabular}
{|c|c|c|c|}
\hline  & $f_i(0)$ & $a_1$ & $a_2$  \\
\hline  $f_1$ & $0.87$ & $-2.45$ & $4.06$  \\
\hline  $f_2$ & $0.21$ & $-3.14$ & $3.95$  \\
\hline  $g_1$ & $0.14$ & $-0.83$ & $1.55$  \\
\hline  $g_2$ & $0.33$ & $-2.45$ & $2.64$  \\
\hline
\end{tabular}
\end{center} \label{weakfit}
\end{table}

\begin{figure}
\begin{minipage}{7cm}
\epsfxsize=6cm \centerline{\epsffile{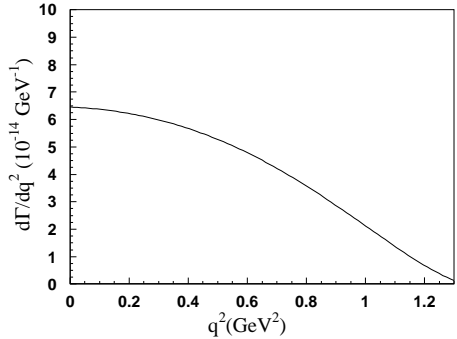}}
\end{minipage}
\caption{\quad $q^2$ dependence of the differential decay rate of $\Xi_c\rightarrow \Xi e^+\bar\nu_e$.}\label{difdecay}
\end{figure}

With Eq. (\ref{decayrate}) and the dipole formula (\ref{weakformulafit}), we can give the $q^2$-dependent differential decay rate, which is shown in Fig.
\ref{difdecay}. In the analysis the following parameters are central values from PDG: $|V_{cs}|=1.04$ and $G_F=1.166 \times 10^{-5}\;\mbox{GeV}^{-2}$. The total
decay width can be obtained by integrating out $q^2$ over the whole kinematical region $0\leq q^2\leq (M_{\Xi_c}-M_\Xi)^2$. The final estimation of the total decay
width is $\Gamma=8.73\times10^{-14}\;\mbox{GeV}$. For a comparison in detail, we first turn to results from the experiments. As there are no absolute branching
fractions available in experiments so far, we only consider the relative branching ratios herein. In Ref. \cite{Alexander}, the relative branching ratios of
$B(\Xi_c^+\rightarrow \Xi^-\pi^+\pi^+)/B(\Xi_c^+\rightarrow\Xi^0 e\nu)$ and $B(\Xi_c^0\rightarrow \Xi^-\pi^+)/B(\Xi_c^0\rightarrow\Xi^- e\nu)$ have been measured.
Making use of the up bounds of the channels $B(\Xi_c^+\rightarrow \Xi^-\pi^+\pi^+)\leq2.1\times10^{-2}$ and $B(\Xi_c^0\rightarrow \Xi^-\pi^+)\leq4.3\times10^{-3}$,
we give our estimations: $B(\Xi_c^+\rightarrow\Xi^0 e\nu)/B(\Xi_c^+\rightarrow \Xi^-\pi^+\pi^+)=2.7$ and $B(\Xi_c^0\rightarrow\Xi^- e\nu)/B(\Xi_c^0\rightarrow
\Xi^-\pi^+)=3.4$, which agree with the values provided by PDG \cite{PDG} $B(\Xi_c^+\rightarrow\Xi^0 e\nu)/B(\Xi_c^+\rightarrow \Xi^-\pi^+\pi^+)=2.3$ and
$B(\Xi_c^0\rightarrow\Xi^- e\nu)/B(\Xi_c^0\rightarrow \Xi^-\pi^+)=3.1$. In addition, we present theoretical results from various models in Table \ref{compare}. It
can be seen from the table that our estimation is acceptable.

\begin{table}
\caption{Decay widths from various models in units of $10^{10} s^{-1}$.}
\begin{center}
\begin{tabular}
{|c|c|c|c|c|c|c|}
\hline  & \cite{Perez} & \cite{Singleton} & \cite{Hussain} & \cite{Chy} & \cite{Ivanov} & this work  \\
\hline  $\Xi_c\rightarrow \Xi e^+\nu_e$ & $28.8(18.1)$ & $8.5$ & $8.55$ & $7.4$ & $8.16$ & $13.26$  \\
\hline
\end{tabular}
\end{center} \label{compare}
\end{table}

\section{Summary}\label{sec:sum}
In this paper, the $\Xi$ baryon distribution amplitudes are investigated up to twist six based on the conformal symmetry. It is found that $14$ independent DAs are
needed to describe the structure of the baryon. In the calculations, the DAs are expanded by the conformal partial waves, and the nonperturbative parameters are
determined in the QCD sum rule approach. The calculation on the conformal expansion of the DAs is to the leading order conformal spin accuracy.

As applications, the electromagnetic form factors of $\Xi$ are investigated in the range $1\; \mbox{GeV}^2\leq Q^2\leq 7\; \mbox{GeV}^2$ with the aid of the DAs
obtained. Our calculations show that the magnetic form factor can be well described by the dipole formula. By fitting the result with the dipole law, the magnetic
moments of the baryons are estimated to be $\mu_{\Xi^0}=-(1.92\pm 0.34)\mu_N$ and $\mu_{\Xi^-}=-(1.19\pm 0.03)\mu_N$. In comparison with the values given by PDG
\cite{PDG}, our results are larger in absolute values. This shows that our calculations need more detailed information on the DAs, which may come from higher order
conformal spin contributions, and at the same time the choice of the interpolating currents may also affect the estimations to some extent
\cite{Lenz,Aliev,Ioffeff}.

We also study the semileptonic weak decay $\Xi_c\rightarrow \Xi e^+\bar\nu_e$. The weak transition form factors are calculated within LCSR method. Our estimation of
the decay width is $\Gamma=8.73\times10^{-14}\;\mbox{GeV}$. We give premature estimations of the relative branching ratios, which are in accordance with the present
experimental data. More experiments are expected to test the calculations and give us more information.

\acknowledgments  This work was supported in part by the National
Natural Science Foundation of China under Contract No.10675167.
\appendix
\section*{Appendix}
In the appendix we give the explicit expressions of the $\Xi$ baryon
DAs. As to the definition in (\ref{da-deftwist}), our results are
listed in this section.

Twist-$3$ distribution amplitudes of $\Xi$ are:
\begin{eqnarray}
V_1(x_i)&=&120x_1x_2x_3\phi_3^0,\hspace{2.5cm}A_1(x_i)=0,\nonumber\\
T_1(x_i)&=&120x_1x_2x_3\phi_3^{'0}.
\end{eqnarray}
Twist-$4$ distribution amplitudes are:
\begin{eqnarray}
S_1(x_i)&=&6(x_2-x_1)x_3(\xi_4^0+\xi_4^{'0}),\hspace{1.6cm}P_1(x_i)=6(x_2-x_1)x_3(\xi_4^0-\xi_4^{'0}),\nonumber\\
V_2(x_i)&=&24x_1x_2\phi_4^0,\hspace{3.9cm}A_2(x_i)=0,\nonumber\\
V_3(x_i)&=&12x_3(1-x_3)\psi_4^0,\hspace{2.8cm}A_3(x_i)=-12x_3(x_1-x_2)\psi_4^0,\nonumber\\
T_2(x_i)&=&24x_1x_2\phi_4^{'0},\hspace{3.9cm}T_3(x_i)=6x_3(1-x_3)(\xi_4^0+\xi_4^{'0}),\nonumber\\
T_7(x_i)&=&6x_3(1-x_3)(\xi_4^{'0}-\xi_4^0).
\end{eqnarray}
Twist-$5$ distribution amplitudes are:
\begin{eqnarray}
S_2(x_i)&=&\frac32(x_1-x_2)(\xi_5^0+\xi_5^{'0}),\hspace{1.5cm}P_2(x_i)=\frac32(x_1-x_2)(\xi_5^0-\xi_5^{'0}),\nonumber\\
V_4(x_i)&=&3(1-x_3)\psi_5^0,\hspace{2.95cm}A_4(x_i)=3(x_1-x_2)\psi_5^0,\nonumber\\
V_5(x_i)&=&6x_3\phi_5^0,\hspace{4.05cm}A_5(x_i)=0,\nonumber\\
T_4(x_i)&=&-\frac32(x_1+x_2)(\xi_5^{'0}+\xi_5^0),\hspace{1.25cm}T_5(x_i)=6x_3\phi_5^{'0},\nonumber\\
T_8(x_i)&=&\frac32(x_1+x_2)(\xi_5^{'0}-\xi_5^0).
\end{eqnarray}
And finally twist-$6$ distribution amplitudes are:
\begin{eqnarray}
V_6(x_i)&=&2\phi_6^0,\hspace{2.5cm}A_6(x_i)=0,\nonumber\\
T_6(x_i)&=&2\phi_6^{'0}.
\end{eqnarray}


\end{document}